\UseRawInputEncoding

\documentclass[10pt,journal,compsoc]{IEEEtran}
\ifCLASSOPTIONcompsoc
  \usepackage[nocompress]{cite}
\else
  \usepackage{cite}
\fi
\ifCLASSINFOpdf
\else
\fi
\usepackage[dvipsnames]{xcolor}
\usepackage{array}
\usepackage{wrapfig}
\usepackage{multirow}
\usepackage{tabularx}
\usepackage{caption}
\usepackage{subcaption}
\usepackage{booktabs}
\usepackage{setspace}
\usepackage{tikz}
\usepackage{amssymb}
\usepackage{booktabs}
\usepackage{enumitem}
\usepackage{multirow}
\usepackage{longtable}
\usepackage{changepage}
\usepackage{hyperref}
\hypersetup{
    colorlinks=true,
    linkcolor=black,
    filecolor=black,      
    urlcolor=black,
    citecolor=black,
}

\usepackage{graphicx}

\usepackage{fancybox}

\begin{document}
%
\title{The Effects of Human Aspects on the Requirements Engineering Process: A Systematic Literature Review}
%
%
%
%

\author{Dulaji~Hidellaarachchi,
        John~Grundy,
        Rashina~Hoda,
        Kashumi~Madampe
\IEEEcompsocitemizethanks{\IEEEcompsocthanksitem Hidellaarachchi, Grundy, Hoda and Madampe are with Faculty of IT, Monash University, Melbourne, Australia \protect\\
Contact E-mail: john.grundy@monash.edu}
\thanks{Manuscript received August 2020}}

%
%

\markboth{Submitted to IEEE Transactions on Software Engineering}%
{Hidellaarachchi \MakeLowercase{\textit{et al.}}: Effects of Human Aspects on RE Process}
%



\IEEEtitleabstractindextext{%
\begin{abstract}
Requirements Engineering (RE) requires the collaboration of various roles in SE, such as requirements engineers, stakeholders and other developers, and it is thus a very highly human dependent process in software engineering (SE). Identifying how \textit{``human aspects"} -- such as personality, motivation, emotions, communication, gender, culture and geographic distribution -- might impact on the RE process would assist us in better supporting successful RE. The main objective of this paper is to systematically review primary studies that have investigated the effects of various human aspects on the RE process. We wanted to identify if any critical human aspects have been found, and what might be the relationships between different human aspects impacting the RE process. A systematic literature review (SLR) was conducted and identified 474 initial primary research studies. These were eventually filtered down to 74 relevant, high-quality primary studies. No primary study to date was found to focus on identifying what are the most influential human aspects on the RE process. Among the studied human aspects, the effects of communication have been considered in many studies of RE. Other human aspects such as personality, motivation and gender have mainly been investigated to date in relation to more general SE studies that include RE as one phase. Findings show that studying more than one human aspect together is beneficial, as this reveals relationships between various human aspects and how they together impact the RE process. However, the majority of these studied combinations of human aspects are unique. From 56.8\% of studies that identified the effects of human aspects on RE, 40.5\% identified the positive impact, 30.9\% negative, 26.2\% identified both impacts whereas 2.3\% mentioned that there was no impact. This implies that a variety of human aspects positively or negatively affects the RE process and a well-defined theoretical analysis on the effects of different human aspects on RE remains to be defined and practically evaluated. The findings of this SLR help researchers who are investigating the impact of various human aspects on the RE process by identifying well-studied research areas, and highlight new areas that should be focused on in future research.
\end{abstract}

}

\maketitle

\IEEEdisplaynontitleabstractindextext

%
\IEEEpeerreviewmaketitle

\IEEEraisesectionheading{\section{Introduction}\label{sec:introduction}}

%
%
%
%

\IEEEPARstart{R}equirements Engineering (RE) is a critical phase of software engineering (SE) where requirements are collected from various sources and are used to define ``what a system should do" vs ``how it should do it". RE activities include eliciting, analysing, documenting, validating and maintaining software requirements \cite{RN2732} \cite{RN2973}. RE is considered to be one of the most important and challenging parts of SE, as it impacts on every stage of the software development process\cite{RN2731}. 
Since SE significantly depends on the performance of the software teams, undertaking RE has become one of the critical responsibilities that software teams must give their attention. A capable software team is thus essential for improving the effectiveness of the  SE process including RE. There needs to be an effective collaboration of individuals with appropriate technical skills and understanding of human issues\cite{RN2733}. 
Software developers, including Requirements Engineers, must work effectively together and with their stakeholders.  

Different \emph{human aspects} will thus impact the RE and SE processes\cite{RN2970}. As RE is a socio-technical activity, it is vital to identify how these various human aspects affect RE, for better or worse, and be able to leverage them to improve the RE process. Identification of the effect of different individual human aspects and combinations of different human aspects in SE activities is  an emerging area of study where researchers are paying more attention. Researchers have investigated human aspects such as \emph{personality}\cite{RN2956}\cite{RN2568}, \emph{emotions}\cite{RN2726}, \emph{motivation}\cite{RN2441}\cite{RN2618}, \emph{gender}\cite{RN2668}, \emph{culture}\cite{RN2593}, \emph{communication issues}\cite{RN2938}\cite{RN2383}, \emph{human errors}\cite{RN2967}, \emph{attitude}\cite{10.1145/1370114.1370127}, \emph{team climate}\cite{RN2712} and others in various SE contexts, where these have sometimes become make-or-break issues that affect many software projects\cite{RN2971}. However, human aspects and their effect on the RE process is still an area that has had relatively limited attention.

\par In this paper we wanted to systematically analyse work done to date that has considered and tried to identify the key effects of human aspects on RE. We focus on RE as it is an inherently and necessarily social process that involves critical contributions of diverse teams and individuals. Furthermore, failures in the RE process will potentially lead to systematic failures in the products that are produced as a result \cite{RN2972}. By conducting this novel Systematic Literature Review (SLR) we aimed to identify what effects various human aspects on the RE process have been studied and found to date. First, we developed an SLR protocol to find and analyse primary studies investigating the effect of diverse human aspects on the RE process, following Kitchenham and Charters' guidelines\cite{RN2753} and Kitchenhams' procedures\cite{RN2754}. After searching and filtering, we found 74 high quality primary studies and extracted information from them. We analysed the range of studies conducted, the different human aspects in RE that they have investigated, evaluated the methodologies these studies used, their solutions, and whether they had been conducted in industry and/or academia. We identified a number of human aspects and combinations that have been demonstrated to impact the RE process. We also identified a number of under-researched or non-researched human aspects and combinations in regards to their impact on RE. We present these and other research gaps identified from the primary studies to suggest areas for further investigation in the RE field related to human aspects.The main contributions of this research are as follows:
\begin{itemize}
    \item A single source of collated information on research into human aspects impact on the RE process
    \item A guidance for IT professionals,  software teams and stakeholders as well as academic and industry researchers who want to better understand the impact of diverse human aspects on the RE process.
    \item A set of recommendations for future research into the impact of human aspects on RE
\end{itemize}
The rest of this paper is organized as follows. Section \ref{section 2} describes key related work in the area of the effect of human aspects on SE and RE processes. Section \ref{section 3} presents our systematic literature review (SLR) research methodology used, and section \ref{section 4} presents and discusses the key findings from this SLR. Finally, section \ref{section 5} discusses key future research recommendations, and section \ref{section 6} concludes this paper. 

\section{Related Work} \label{section 2}
Although the interaction between humans and computers has a long history, investigating the relationship between human aspects and SE has become an emerging area of considerable research in recent years \cite{RN2587}. Much of software engineering is in many aspects a human-centred activity \cite{john2005human}. \textbf{Requirements Engineering} (RE) is arguably the most human-centric activity in SE, requiring people involved in it needing to work closely and effectively with diverse stakeholders, software development team members, and other requirements engineers\cite{RN2952,RN2445}. 
\subsection{Human aspects in Software Engineering} \label{section 2.1}
 Various human aspects have been shown to have impact on different stages in the SE process. The majority of these studies to date have focused on impacts during software design and implementation. According to the SLR conducted by \cite{RN2966}, these development stages have been focused on in 94\% of their identified papers. They also claimed that, despite the impact of human aspects in SE, researchers have still not paid enough attention to this area.

A number of SLRs have been conducted targeting the identification of various \textbf{human aspects of software engineers}, such as \textit{motivation, creativity, personality, behaviour, gender equity, human values, self-management barriers and self-compassion}. Cruz et al. \cite{RN2702} in their systematic mapping study reviewed  research on \textbf{personality} in SE. They analysed many published empirical and theoretical studies related to role of personality on different aspects of SE. Based on their findings, pair programming, education, software engineers' personality characteristics, and team effectiveness related to personality were identified as the most focused on areas. They identified that the number of articles related to personality and SE has significantly increased after 2002. An SLR was conducted by Barroso et al. \cite{RN2956} that focused on the influence of human personality in SE. In this study, they evaluated personality models and tests applied by SE researches and identified how human personality influences software engineers' work.  The majority of the studies  focused on software designer and coder personalities.

Soomro et al. \cite{RN2712} conducted an SLR on the effect of software engineers' personality and how it is associated with \textbf{team climate} and \textbf{team performance}. Their findings revealed that there is a relationship between software engineers' personality and team performance, without considering team climate. Their study also revealed that software project team characteristics have a significant impact on software team performance, and diverse team climate compositions have been discussed mainly in terms of organizational behaviour and social science domains, but not much in an SE context.

Another SLR related to various \textbf{human behavioural} aspects in SE was conducted by Lenberg et al. \cite{RN2968}. This aimed to create a common platform for future research in the area. They suggested a new research area as \textit{behavioural software engineering} (BSE) and presented a definition of BSE as \emph{"the study of behavioral and social aspects of SE activities performed by individuals, groups or organizations"}. The results of their research indicated that BSE is an emerging research area where the majority of researches are based on software engineers, teams or organizations in general. They found that specific phases or activities in SE have not yet been frequently considered. Moreover, they identified that there is an imbalance of studies that focused on human aspects, as most of the studies considered \textit{communication, personality} and \textit{job satisfaction} related to software engineers. They suggest that researchers should explore more human aspects and consider their impact on a wider range of SE activities. 

Hall et al. \cite{RN2572} conducted a systematic review to identify theory use in studies investigating the \textbf{motivations} of software engineers. By analysing 92 studies related to  motivation in SE, they found that many studies have focused on motivation of software engineers, but not explicitly underpinned by existing  motivational theories. However, the findings of the reviewed primary studies showed a clear relationship with these theories. Sach et al. \cite{RN2582} also focused on motivational factors in software development where 23 software practitioners were engaged for a workshop on motivation and collected data to investigate motivational factors that effect on their software development practices. Based on their results, they claim the \emph{people} factor is the most commonly listed motivational factor for software practitioners, compared to other factors, such as financial, autonomy, and learning. In \cite{RN2713}, an empirical study was conducted to investigate on how software testers can be motivated. Semi-structured in-depth interviews were conducted with 36 practitioners in 12 software organizations in Norway. Set of motivational and demotivational factors influencing software testing personnel were identified and proposed that combining testing responsibilities with variety of tasks engagement increase the satisfaction of testers which eventually increase their motivation.

\textbf{Gender} is another human aspect that is emerging in importance in SE research and practice. In \cite{RN2668}, a case study was used to investigate  gender equality in a national software academy (NSA). This research tried to identify the experience of gender equality over three years in NSA and discussed  measures to be implemented in future research to raise awareness and reduce the gender gap among all levels at the NSA. In the study \cite{RN2682}, \textbf{human values} were measured related to SE where they investigated the influence of human values in the software production decision-making process. The researchers considered human values as a mental representation and investigated them based on three levels -- system level, personal level and instantiation level. Three human values prototypes were identified for software practitioner, the \textit{intrinsically-driven socially-concerned practitioner, the autonomous nonconforming risk-taker} and \textit{the fun-loving extrinsically-driven practitioner}. The researchers claimed that this approach should be used more widely so that researchers don't miss values in future research. A systematic mapping study was conducted on \textbf{soft skills} in software engineering by \cite{RN1592}, aiming to identify what soft skills are considered relevant to the practice of software engineering. They focused on categorizing 30 soft skill categories based on the definitions taken from those primary studies. They identified \textit{communication skills, team work, analytical skills, organizational skills, and interpersonal skills} as most studied soft skills in software engineering. In \cite{RN1593}, a systematic review was conducted aiming to identify soft skills for the IT project success. Their research also revealed list of soft skills that impact the success of the project and among those, \textit{communication, leadership, conflict management, thinking, Innovativeness, change orientation, negotiation, motivation and problem solving} were identified as most mentioned soft skills in the literature. 

A systematic mapping of \textbf{human cognitive biases} in SE showed that software engineers are susceptible to a range of biased decision making at different phases of development \cite{mohanani2018cognitive}. They highlight a lack of good mitigation techniques and limited theoretical foundations for interpreting biases in this area. They suggest some techniques to mitigate bias, but also highlight the need for further studies of biased human decision making in SE, including in RE.

\subsection{Human aspects in Requirements Engineering} \label{section 2.2}
In terms of studies that focused on the effects of human aspects on the RE process in particular, \cite{RN2952} and \cite{RN2486} focused on effective \textbf{communication} as a critical success factor during requirements elicitation. However these studies were limited to global software development (GSD) and identified that effective  communication plays a significant role in requirements elicitation specifically for GSD teams. It was found that \textit{geographical distribution}, \textit{time zone}, \textit{cultural diversity} and \textit{physical differences} were reasons for miscommunication when conducting requirements elicitation in GSD. Their analysis also indicates that lack of effective communication, lack of knowledge sharing and awareness, lack of collaboration and organizational change are common critical challenges related to the RE process in GSD. Khan and Akbar \cite{RN2215} performed an SLR and an empirical investigation on \textbf{motivation factors} for the requirements change management process in GSD. They explored the motivators that contribute to requirements change management by extracting 25 motivators, and developed taxonomies of identified motivators such as accountability,  clear change management strategy, overseas site's response, and effective change management leadership. 

Anu et al. \cite{RN2967} conducted a systematic study on \textbf{human error} research, focusing on both the SE and psychology literature to identify and classify human errors that occur during the RE process. A human error taxonomy (HET) was proposed that is based on a standard human error taxonomy from cognitive psychology. This can be used to identify the most common errors made in the RE process and aims to help improve the quality of the resultant software. Although there are many systematic studies that have focused on the effects of various human aspects on RE, the majority of the studies are limited to studying the particular RE process issues that occur in the GSD domain. Other studies have focused on only one particular human aspect or one phase of the RE process e.g. the elicitation phase. There is lack of a systematic study that focuses on identifying the effects of  human aspects considering all RE process phases.

\cite{RN2430} focused on classifying effective personalities for web development during requirements elicitation. Their research revealed that there is a relationship between human personalities and RE in web development and a need more research that considers more human aspects and their impact on the RE process. Aldave et al. \cite{RN2969} conducted an SLR to identify the influence of \textbf{creativity} on requirements elicitation within agile software development. They found that enhancing creativity in requirements elicitation can be implemented successfully in agile based software projects, specifically user interface development projects. Moreover, they identified that creativity is an important aspect in SE which brings innovation to the project. Despite their findings, they say that more research is required to understand the influence of creativity in RE.

Most of these systematic and empirical studies have  focused on various human aspects related to SE in general, or predominantly on design and development phase of SE, or specifically on agile teams and GSD contexts. The studies that focused on RE have been mainly limited to GSD, web development or a particular phase in the RE  process, usually requirements elicitation. Moreover, the majority of the studies have focused on identifying the effect of one human aspect on SE or RE processes. Cheng and Atlee \cite{RN2365} discussed current and future research directions in RE. They claim that identifying human behaviour in RE is an open and very challenging problem and it has become a key emerging area for RE researchers. As longer term actions that would help the RE community of research, they state that RE researchers should think beyond current RE and SE knowledge and collaborate with other disciplines to improve the RE process, including identification of better methods to model human behaviours in RE. This highlights the need for a systematic review that identifies and analyses  primary studies focusing on a range of human aspects across the whole RE process. 

\section{Research Methodology} \label{section 3}

This Systematic Literature Review (SLR) was conducted to evaluate and synthesize existing primary research studies on the effect of human aspects on Requirements Engineering (RE).  The SLR is aimed to provide an in-depth analysis of the research to date in this field relevant to specific research questions.  We followed the well-defined methodology described by Kitchenham and Charters' guideline\cite{RN2753} and Kitchenhams' procedures\cite{RN2754} in conducting this SLR, to make it as unbiased and repeatable in assessing all possible evidence from published primary studies. Our findings will be beneficial for software engineering researchers by providing an analysis of existing information on study methodologies used, solutions claimed,  their usage in  industry or academia, and by identifying research gaps in order to suggest key areas for further investigation in the RE field related to human aspects. 

The first author defined a detailed review protocol, conducted detailed searches, filtered the studies, and carried out data extraction and analysis under the close supervision of the second and third authors, both very  experienced in conducting SLRs in SE. To synthesize the extracted data from the final 74 studies, listed in Appendix B, we performed a meta-analysis technique \cite{RN1595}. This can be described as an analysis of a large collection of structured, extracted data from individual studies for the purpose of integrating and summarising their key findings, a well-accepted approach in SLRs.

\subsection{Research Questions}
Initially, the set of research questions (RQs) were developed by following the approach of Petticrew and Roberts \cite{RN2960}: intervention, population, outcomes of interest, and context within which the intervention is delivered. This approach is explained in the Kitchenham and Charters' guideline\cite{RN2753} as PICOC (population, intervention, comparison, outcomes and context). The following RQs were developed with the assistance of the PICOC in Table \ref{TABLE 1: PICOC}. 

\par \textbf{RQ1. What is the motivation behind each primary study on identifying the effects of human aspects in requirements engineering?} -- This research question examines the main goals, objectives and motivation behind identifying the effect of human aspects on the RE process in each primary study. We also looked at what the target system domain and whether the study was in an academic or industry setting. 

\textbf{RQ2. What is the current status of research studies on the effect of human aspects in requirements engineering?} -- This research question focuses on identifying  the human aspects that have been investigated relating to the RE process to date. We also examine the kind of methodologies used by the researchers to identify the effect of human aspects on RE, types of existing domain models used by the researchers, what solutions have been presented to address the effect of human aspects on the RE process and how they have performed the evaluation of the solutions presented. Furthermore, we attempt to identify the key limitations and future work areas of each primary study. 

 \textbf{RQ3. What RE phases are most impacted by human aspects, and what are the relationships between different human aspects that affect these RE phases?} -- This research question investigates which RE phases are most impacted by human factors. It also examines how these human aspects impacting RE are related to one another, and how we might categorize the most important (combinations of) human aspects that are important to consider for RE.

 \textbf{RQ4. How do the identified human aspects affect the RE Process?} -- This research question asks what have been identified as the key effects on RE phases by different human aspects. If the effect is positive, then what are the benefits of considering those human aspects impacts on RE. If the effect is negative, then what are the approaches that have been used to mitigate the negative effects. 

\begin{table}[]
\centering
\caption{PICOC for Research Questions}
\label{TABLE 1: PICOC}
\resizebox{\columnwidth}{!}{%
\begin{tabular}{@{}ll@{}}
\toprule
\footnotesize
\textbf{Population}   & \begin{tabular}[c]{@{}l@{}}Requirements Engineers/  Business Analysts/ \\   Software Practitioners involved in RE      \\ \end{tabular}                                                                                                                                                   \\
\textbf{Intervention} & Human aspects on Requirements Engineering (RE)                                                                                                                                                            \\
\textbf{Comparison}   & N/A                                                                                                                                                                                                       \\
\textbf{Outcomes}     & \begin{tabular}[c]{@{}l@{}}Effects of Human aspects on RE \\  Relationship among Human aspects    \\ \end{tabular}                                                                                                                                 \\
\textbf{Context}      & \begin{tabular}[c]{@{}l@{}}Requirements Engineering/ Requirements Elicitation/ \\ Requirements Specification/ Requirements Analysis/ \\ Requirements Validation/ Requirements Management/ SE\end{tabular} \\ \bottomrule
\end{tabular}%
}
\end{table}

\begin{table}[t]
\centering
\caption{Major search terms}
\label{TABLE 2: Major Terms}
\resizebox{\columnwidth}{!}{%
\begin{tabular}{@{}ll@{}}
\toprule
\footnotesize
\textbf{Intervention} & \begin{tabular}[c]{@{}l@{}}Human aspects/ Human-centric issues/ \\ Human factors on Requirements Engineering (RE)\end{tabular} \\
\textbf{Outcomes}     & Effects of Human aspects on RE                                                                                                 \\
\textbf{Context}      & Requirements Engineering                 \\ \bottomrule                                                                                     
\end{tabular}%
}
\end{table}

\begin{table}[]
\centering
\caption{Alternative search terms}
\label{TABLE 3: Alternative terms}
\resizebox{\columnwidth}{!}{%
\begin{tabular}{@{}ll@{}}
\toprule
\textbf{\begin{tabular}[c]{@{}l@{}}Humancentric issues/ \\ Human factors\end{tabular}} & \begin{tabular}[c]{@{}l@{}}Personality/ Culture/ Emotions/ Age/ Human Values/ \\ Creativity/ Gender/ Communication issues/ \\ Physical issues/ Psychological issues\end{tabular}                  \\
\textbf{\begin{tabular}[c]{@{}l@{}}Requirements \\ Engineer\end{tabular}}              & \begin{tabular}[c]{@{}l@{}}Requirements Practitioners/ Business Analysts/ \\ Software Engineer\end{tabular}                                                                                       \\
\textbf{\begin{tabular}[c]{@{}l@{}}Requirements \\ Engineering\end{tabular}}           & \begin{tabular}[c]{@{}l@{}}Requirements Elicitation/ Requirements Specification/ \\ Requirements Analysis/ Requirements Validation/ \\ Requirements Management/ Software Engineering\end{tabular}
\\ \bottomrule
\end{tabular}%
}
\end{table}

\begin{table}[t]
\centering
\caption{Construction of search strings with OR}
\label{TABLE 4: Constructed search strings}
\resizebox{\columnwidth}{!}{%
\begin{tabular}{@{}cl@{}}
\toprule
\footnotesize
1 & \begin{tabular}[c]{@{}l@{}}Human-centric issues OR Human factors OR Personality OR \\ Culture OR Emotions OR Age OR Human Values OR \\ Creativity OR Gender OR Communication issues OR \\ Physical issues OR Psychological issues\end{tabular} \\
2 & \begin{tabular}[c]{@{}l@{}}Requirements Engineering OR Requirements Elicitation OR \\ Requirements Specification OR Requirements Analysis OR \\ Requirements Validation OR Requirements Management OR \\ Software Engineering\end{tabular}     \\
3 & \begin{tabular}[c]{@{}l@{}}Requirements Engineer OR Requirements practitioners OR \\ Business Analysts OR Software Engineer\end{tabular}                                            \\ \bottomrule                                                          
\end{tabular}%
}
\end{table}

 \begin{figure*}[t]
 \centering
  \includegraphics[width=0.7\linewidth,height=6cm]{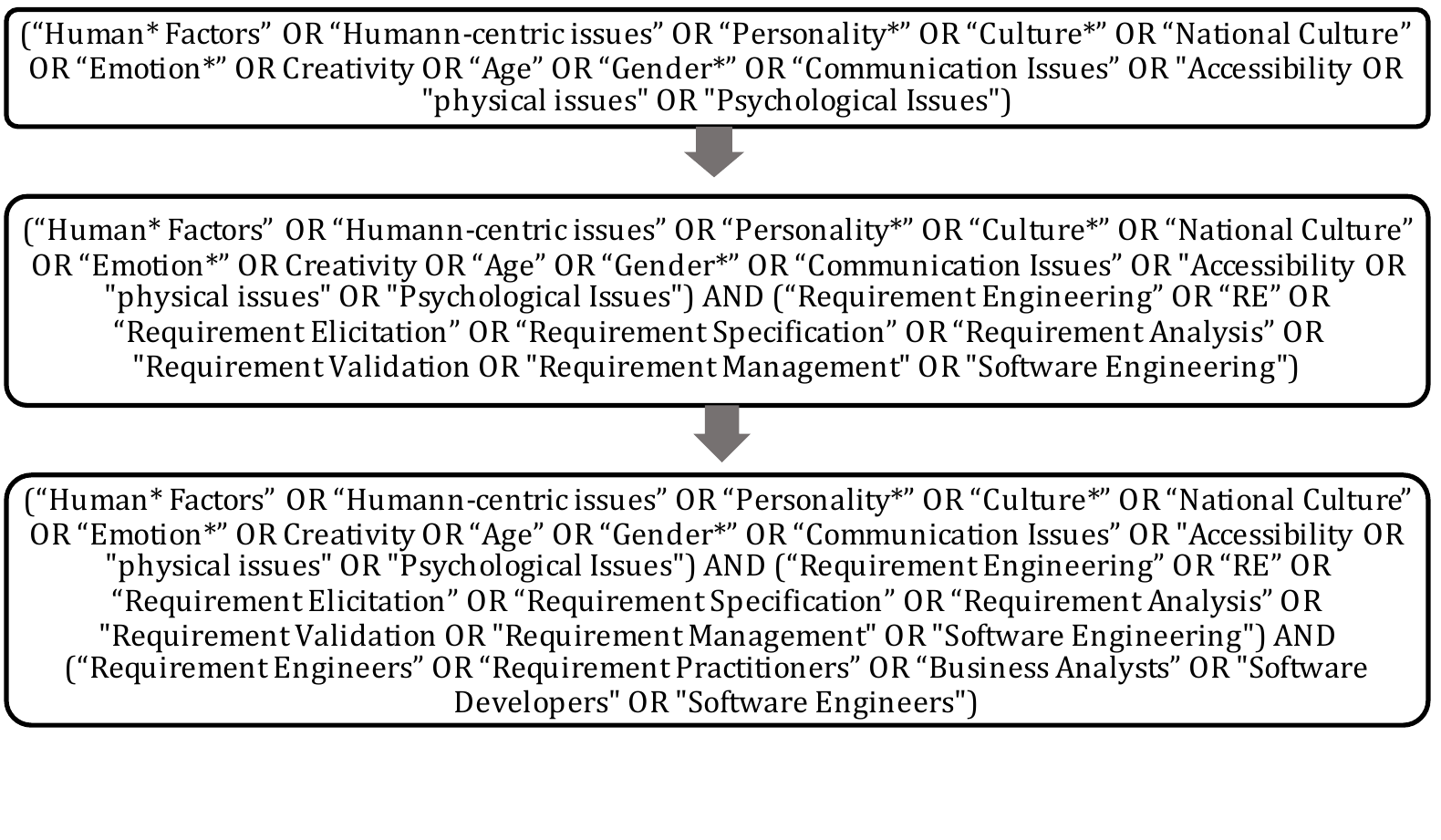}
  \caption{Formulation of search string}
  \label{Search_String}
\end{figure*}
 \subsection{Identifying the relevant literature}
 To search and identify relevant primary studies for this SLR, we defined search terms as shown in Table \ref{TABLE 2: Major Terms} and Table \ref{TABLE 3: Alternative terms} and followed a systematic search strategy. 
 
 \subsubsection{Search strategy for primary studies}
\par First, search queries were developed by selecting key search terms from PICOC (Table \ref{TABLE 1: PICOC} and Table \ref{TABLE 2: Major Terms}). To obtain more relevant primary studies, alternative search terms were also used aligning with the main concept (Table \ref{TABLE 3: Alternative terms}). Based on all the identified main and alternative search terms, several search strings were developed and the execution of the search strings were carried out on four (4) online databases (Table \ref{TABLE 6: Search strings}). When developing the final search string, the search terms were linked mostly with Boolean AND and OR operators and one proximity operator, NEAR. The AND operator was used to concatenate the key terms, OR to link the synonyms and NEAR to find primary studies where the terms joined within a specified number of words of each other (Table \ref{TABLE 2: Major Terms} , Table \ref{TABLE 3: Alternative terms} and Table \ref{TABLE 4: Constructed search strings}). Several search iterations were carried out to find out the best search strings for each database and during the search process, we have used several search tips (eg: wildcards, stemming) as instructed in digital library help sections. The best search strings for each online database were selected based on the most relevant primary studies given by the search string. We have not selected any particular time range for the search as we wanted to examine the distribution of all the identified primary studies. Considering the above strategies, we have formulated a primary search string for the SLR as shown in Figure \ref{Search_String}.

 \subsubsection{Primary and Secondary search process}
To identify relevant research papers for the SLR, the searching procedure was conducted in two ways, automatic and manual. Automatic search was performed by using scientific databases' search engines and manual by scanning the list of references of primary studies gained from automatic search (Table \ref{TABLE 5: Refined areas}).
The scientific databases in Table \ref{TABLE 5: Refined areas} were selected based on having most papers in computer science and SE studies. 
We also did manual searching via backward and forward snowballing using the retrieved primary studies, which helped us find other relevant studies for our SLR as suggested by Achimugu et al. \cite{RN2961}. We selected 8 highly related papers out of 16 papers we collected from this manual search (Table \ref{tab:paper_count}).

\subsubsection{Refining the search string}
 Due to the limitations of each database, we  refined our formulated search string according to the requirements and setup of the search engines of databases (Table \ref{TABLE 5: Refined areas}). 
Next, to come up with most relevant list of studies, each search string was refined and executed several times. We randomly picked 8-10 papers from each database to verify that the obtained list of studies were the most relevant for our review. The final search strings  used are shown in Table \ref{TABLE 6: Search strings}.
 
\begin{table}[]
\centering
\caption{Refined areas of scientific databases}
\label{TABLE 5: Refined areas}
\resizebox{\columnwidth}{!}{%
\begin{tabular}{@{}lll@{}}
\toprule
\multicolumn{1}{c}{\textbf{Scientific database}}               & \multicolumn{1}{c}{\textbf{Search type}} & \multicolumn{1}{c}{\textbf{Refinement}}                                                                                                                                \\ \midrule
IEEE Xplore                                                    & Command search                           & \begin{tabular}[c]{@{}l@{}}Limited to journals and conferences, \\ used index terms: human factors \\ and software engineering, \\ no specific time range\end{tabular} \\
\begin{tabular}[c]{@{}l@{}}ACM Digital \\ library\end{tabular} & Advanced search                          & \begin{tabular}[c]{@{}l@{}}Limited to journals and conferences, \\ no specific time range\end{tabular}                                                                 \\
Springer                                                       & Advanced search                          & \begin{tabular}[c]{@{}l@{}}Limited to journals and conferences, \\ no specific time range\end{tabular}                                                                 \\
Wiley                                                          & Advanced search                          & \begin{tabular}[c]{@{}l@{}}Limited to journals, \\ no specific time range\end{tabular}                                                                                 \\ \bottomrule
\end{tabular}%
}
\end{table}

 \subsection{Paper selection criteria}
 \subsubsection{Inclusion and exclusion criteria}
 Selection of studies was conducted based on a solid inclusion and exclusion criteria, initially defined when preparing the protocol for this SLR. We used 5 inclusion criteria (Table \ref{TABLE 7: Inclusion criteria}) and 6 exclusion criteria (Table \ref{TABLE 8: Exclusion criteria}) to filter the found papers, to ensure the final papers were inline with our review objectives and research questions. We refined these criteria during search and paper filtering processes to achieve an unbiased set of papers. The finalised set of criteria was applied on all the downloaded full text papers to select the most relevant studies. We did not include any review papers, workshop or magazine articles, or grey literature, in keeping with usual SLR primary studies filtering practice. Though we found few papers including all main keywords of this SLR research -- human aspects, Requirement Engineers and Requirements Engineering Process -- together in one paper, many studies focused on at least one or two human aspects related to specifically RE, or to SE where RE was considered as one phase.

\begin{table}[]
\centering
\caption{List of search strings of each database}
\label{TABLE 6: Search strings}
\resizebox{\columnwidth}{!}{%
\begin{tabular}{@{}lll@{}}
\toprule
\multicolumn{1}{c}{\textbf{Scientific database}}               & \multicolumn{2}{c}{\textbf{Search string}}                                                                                                                                                                                                                                                                                                                                                                                                                                                                                                                                                                                                                     \\ \midrule
IEEE Xplore                                                    & \multicolumn{2}{l}{\begin{tabular}[c]{@{}l@{}}(“Human* Factors” OR “Human-centric issues” OR “Personality*” \\ OR “Culture*” OR “National Culture” OR “Emotion*” OR Creativity \\ OR “Age” OR “Gender*” OR “Communication Issues” OR \\ ”Accessibility OR ”physical issues” OR ”Psychological Issues”) \\ NEAR/10 (“Requirement Engineering” OR “RE” OR \\ “Requirement Elicitation” OR “Requirement Specification” OR \\ “Requirement Analysis” OR ”Requirement Management”) AND \\ (“Requirement Engineers” OR “Requirement Practitioners” OR \\ “Business Analysts” OR ”Software Developers” OR \\ ”Software Engineers”)\end{tabular}}                      \\
\begin{tabular}[c]{@{}l@{}}ACM Digital \\ library\end{tabular} & \multicolumn{2}{l}{\begin{tabular}[c]{@{}l@{}}AllField:((”Human Factors*” OR ”Human values” OR Personality* \\ OR Culture* OR Emotion* OR Age OR creativity OR Gender* OR \\ ”communication issues” OR Accessibility OR ”physical issues” OR \\ ”Psychological Issues”) AND (”Requirement Engineer” OR \\ ”Software Engineer”) AND (”Requirement Engineering” OR RE \\ OR ”Requirement Elicitation” OR ”Requirement Specification” \\ OR ”Requirement Analysis” OR ”Requirement Management” \\ ”Requirement Validation” OR ”Software Engineering”)) ”\\ filter”: Article Type: Research Article\end{tabular}}                                                  \\
Springer                                                       & \multicolumn{2}{l}{\begin{tabular}[c]{@{}l@{}}’(“Human* Factors” OR “Human-centric Issues” OR “Influences” \\ OR “Individual Personality*” OR “Cultural Influence” OR ”Creativity” \\ OR “Emotion*” OR “Age” OR “Gender*” OR “Communication Issues” \\ OR ”Accessibility” OR ”physical issues” OR ”Psychological Issues”) \\ AND (“Requirement Engineering” OR “RE” OR “Requirement Elicitation” \\ OR “Requirement Specification” OR “Requirement Analysis” OR \\ ”Requirement management” OR ”Requirement Validation”) AND \\ (“Requirement Engineers” OR “Requirement Practitioners” OR \\ “Business Analysts”) AND (”Software Engineering”)’\end{tabular}} \\
Wiley                                                          & \multicolumn{2}{l}{\begin{tabular}[c]{@{}l@{}}(“Human* Factors” OR “Human-centric Issues” OR “Personality*” \\ OR “Cultural Aspect” OR “Emotion*” OR “Age” OR “Gender*” \\ OR ”Creativity” OR “Communication Issues” OR ”physical issues” \\ OR ”Psychological Issues”) AND (“Requirement Engineering” OR \\ “RE” OR “Requirement Elicitation” OR “Requirement Specification” \\ OR “Requirement Analysis” OR ”Requirement management” OR \\ ”Requirement Validation”) AND (“Requirement Engineer*” OR \\ “Requirement Practitioners” OR “software engineers”)\end{tabular}}                                                                                   \\ \bottomrule
\end{tabular}%
}
\end{table}

\begin{table}[]
\centering
\caption{Inclusion criteria}
\label{TABLE 7: Inclusion criteria}
\resizebox{\columnwidth}{!}{%
\begin{tabular}{@{}ll@{}}
\toprule
\multicolumn{1}{c}{\textbf{Criterion ID}} & \multicolumn{1}{c}{\textbf{Criterion}}                                                                                                        \\ \midrule
I01                                       & \begin{tabular}[c]{@{}l@{}}Full text papers published as journal or \\ conference Papers that comply with \\ human aspects in RE\end{tabular} \\
I02                                       & Papers that are written in English Language                                                                                                   \\
I03                                       & \begin{tabular}[c]{@{}l@{}}Studies that have been used in academia \\ (Literature references)\end{tabular}                                    \\
I04                                       & \begin{tabular}[c]{@{}l@{}}Papers that are titled as software Engineering/\\ software Engineers, but considered RE phase as well\end{tabular} \\
I05                                       & \begin{tabular}[c]{@{}l@{}}Papers about human-centric issues in software \\ development life cycle including RE\end{tabular}                  \\ \bottomrule
\end{tabular}%
}
\end{table}

\begin{table}[]
\centering
\caption{Exclusion criteria}
\label{TABLE 8: Exclusion criteria}
\resizebox{\columnwidth}{!}{%
\begin{tabular}{@{}ll@{}}
\toprule
\multicolumn{1}{c}{\textbf{Criterion ID}} & \multicolumn{1}{c}{\textbf{Criterion}}                                                                                                                                                                                                   \\ \midrule
E01                                       & \begin{tabular}[c]{@{}l@{}}Workshop articles, books, gray literature \\ (theses, unpublished and incomplete work), \\ posters, secondary or review studies (SLR or SMS), \\ surveys, discussions and keynotes are excluded.\end{tabular} \\
E02                                       & Short papers where page count is less than 4 pages                                                                                                                                                                                       \\
E03                                       & \begin{tabular}[c]{@{}l@{}}Papers with inadequate information to extract \\ (Irrelevant Papers)\end{tabular}                                                                                                                             \\
E04                                       & \begin{tabular}[c]{@{}l@{}}Papers about RE but not discussing about \\ human aspects\end{tabular}                                                                                                                                        \\
E05                                       & \begin{tabular}[c]{@{}l@{}}Papers regarding software engineering/programming/\\ development which are not included RE\end{tabular}                                                                                                       \\
E06                                       & Extended journal article of the same paper                                                                                                                                                                                               \\ \bottomrule
\end{tabular}%
}
\end{table}

\subsubsection{Filtering of the papers}
The filtering process consisted of three screenings as follows:
\begin{itemize}
 \item Initial paper pool: We downloaded 472 potentially relevant papers from the selected scientific databases using our search strings. We then applied our inclusion and exclusion criteria to each. 
  \item First screening: From this initial paper pool, 180 papers were left after using the paper title and the abstract to screen.
  \item Second screening: Further filtering was conducted by reading the title, abstract, conclusion, skimming the introduction, methodology and results. 92  papers were left after this second screening.
  \item Third screening: Further filtering while engaging in the data extraction process. Though we  arrived at most of the relevant studies with the second screening, there were some papers beyond the planned scope which in the end didn't answer our research question-based fields in our data extraction form. For these papers, a third screening was required to decide whether to keep the paper or not, leaving 66 papers.
\end{itemize}

\par After going through this filtering process we obtained 66 papers from our primary database search process, and a further 8 papers from our secondary snowballing search process and filtering. This gave us 74 research papers highly related to our focused area as our final primary studies paper count. The breakdown of the paper count from screening is shown in TABLE \ref{tab:paper_count}.

\begin{table*}[]
\caption{Breakdown of the paper count}
\label{tab:paper_count}
\resizebox{\textwidth}{!}{%
\begin{tabular}{@{}ccccccc@{}}
\toprule
\multicolumn{1}{l}{}                 
& \textbf{Resource name} & \textbf{Initial paper count} & \textbf{Downloaded paper count} 
& \textbf{1st screening} & \textbf{2nd screening} & \textbf{3rd/ Final screening} \\\midrule

\multirow{5}{*}{Primary Search}  & IEEE    & 128   & 127    & 56      & 36    & 27     \\
                                 & ACM DL  & 151   & 151    & 68      & 28    & 19      \\
                                & Springer & 124   & 123    & 34      & 21    & 16      \\
                                & Wiley    & 71    & 71     & 22      & 7    & 4         \\
    
    & \textbf{Count}        & \textbf{474}    & \textbf{472}  & \textbf{180}   & \textbf{92}            & \textbf{66}   \\ \midrule
\multicolumn{1}{l}{Secondary Search} & Snowballing  & 16     & 16      & 16  & 8    & 8                             \\ \midrule
\multicolumn{6}{c}{\textbf{Total final paper count (Primary + Secondary)}} 
& \textbf{74}    \\ \bottomrule 
\end{tabular}%
}
\end{table*}

\par \emph{EndNote library} and \emph{excel sheets} were used to maintain the relevant records of the papers from the initial step to the final screening. In EndNote, separate groups were created to maintain paper lists according to database and separate sheets were maintained in excel with unique colour codes to identify how much a paper is relevant to the SLR. (Eg: Dark green: Highly relevant, Green: Relevant, Yellow: Somewhat relevant, Red: Irrelevant). Based on these colour codes, we could manage which papers required more attention on filtering. 

\subsection{Quality checklist and procedures} \label{3.4}
A quality assessment criteria were used as an approach to evaluate the quality of identified studies. We used two separate ways to check the quality of filtered papers. 

\emph{Publication venues of the final paper list:}
We have searched the ranking of the publication venues of the final paper list to check whether the papers have been published in highly ranked venues. From 74 papers, we were able to search the publication rankings of 69 papers. Based on the CORE 2018 rankings, 13 papers were published in an A* journal or a conference; 22 papers were published in an A ranked journal or a conference; 17 papers were belong to B ranked journals or conferences; and only 6 papers were found which were published in a C ranked conference or journal. This implies that the majority of the papers reporting primary studies analysed belong to top ranked venues. 

\emph{Scoring mechanism with predefined questions related to paper's quality:} We developed a scoring mechanism which consists of five scoring types (low to High) namely very poor (1), inadequate (2), moderate (3), good (4) and excellent (5). Each paper was categorized on the score of 1 to 5 (low to high) by answering the follow questions: 
\begin{enumerate}[label=(\roman*)]
\item Is the paper highly applicable to our SLR research focus?
\item Is there a clear statement of the aim of the research? 
\item Is there a review of key past work? 
\item Is there a clear methodology for the research which aligned with key research questions claimed for the research? 
\item Does the paper provide adequate information regarding the data collection and data analysis of the research? 
 \item Are the findings of the research clearly stated and supported by the research Questions?
 \item Does the paper provide limitations, summary and future work of the research? 
\end{enumerate}

\begin{figure}
  \includegraphics[width=\linewidth]{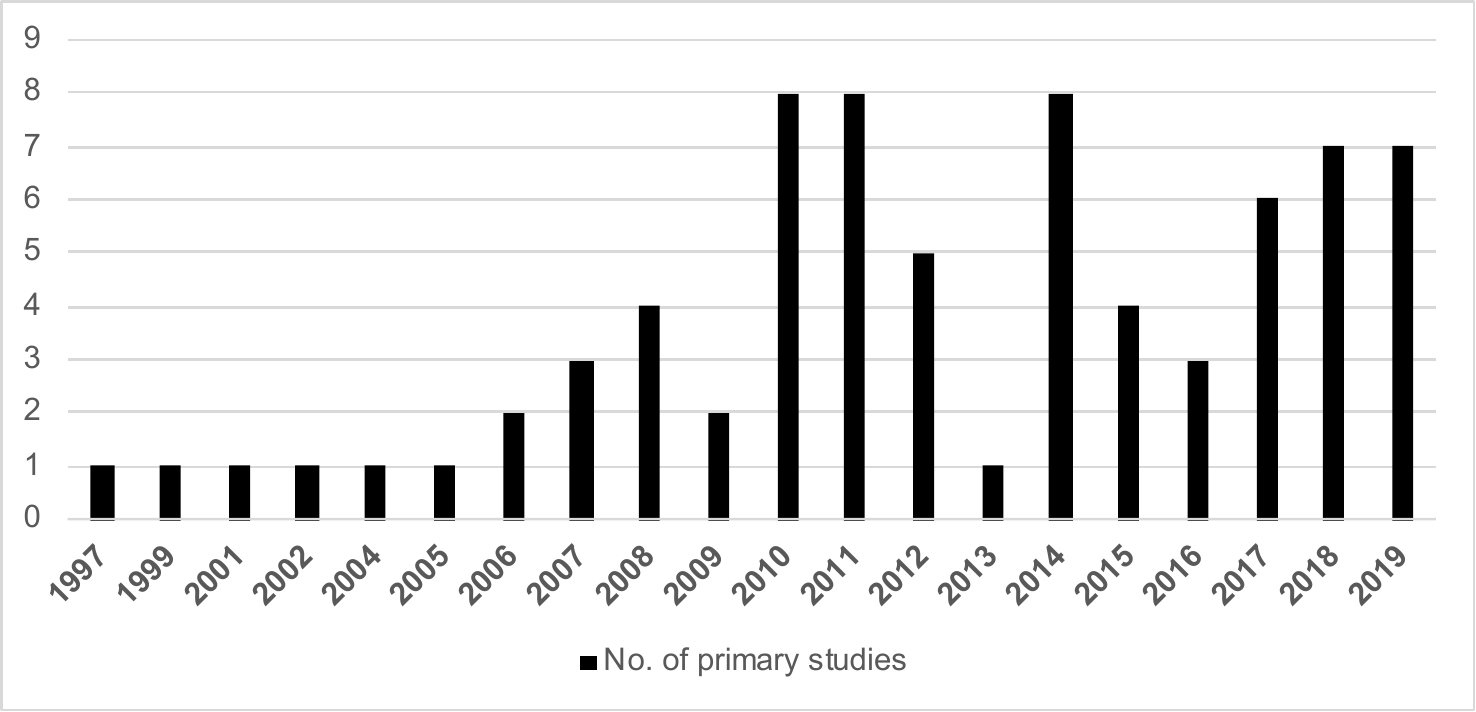}
  \caption{Number of selected primary studies by year}
  \label{PubYear}
\end{figure}

This scoring mechanism was applied for final set of papers where we checked the quality of the filtered studies. We identified 8 low quality studies with average score was less than 3 and filtered these out from the paper set, with final paper count of 74. 
Due to the fluctuation of number of primary studies published overtime we have not included "citation count of each paper" as a quality criteria to be unbiased with papers published in recent years (Figure \ref{PubYear}).

\subsection{Data Extraction Strategy}

To ensure that each research paper was analysed and extracted consistently, a google form was created for the data extraction process with 38 questions, listed in Appendix A. These consist of 17 long answer questions, 9 multiple-choice questions, 6 check-boxes and 6 short answer questions. We grouped this google form into five sections based on the areas focused when extracting data -- general information (paper title, authors and their affiliations, published year and venue); key areas of the study (goal, research questions, subjects used in the studies, focused human aspects and RE phase); methodology used; and key research gaps, research outcomes and future research proposed (proposed solutions, developed model/framework and recommendations). 

Before starting data extraction, the questions and the structure of the google form were fine-tuned three times by choosing papers from each online database and conducting extractions (pilot tests). Selected papers and the google form were then sent to two co-authors to do the same extraction. They  independently extracted the given paper data and we did a comparison to check whether there was any conflicting extracted information. A very high similarity of extracted data was found when using different paper formats and contents. With further discussion and reaching consensus over any disagreement, a final version of the google form was created and the first author carried out the data extraction under the close supervision of second and third authors, with cross-checking of several selected paper data extraction.

%


\section{Findings} \label{section 4}
Data from the final 74 filtered primary studies papers was extracted and contained qualitative, quantitative and mixed data types. Various visualization tools and meta-analysis techniques were used to analyse and present the data (bar chart,pie chart, pivot chart and various tables). The data synthesis spreadsheet can be found \href{http://ieee-dataport.org/documents/data-synthesis-effects-human-aspects-requirements-engineering-process-systematic}{\textbf{here}.}

Of the 74 primary studies included in this SLR, there were 56 conference papers and 18 journal papers, or 76\% and 24\% respectively from the overall paper count. 
All of the selected primary studies were published within the period of 1997- 2019. Figure \ref{PubYear} shows selected primary studies by publication year. We didn't have any included studies in years 1998, 2000 and 2003. From 1997 to 2005, the number of publications per year was low and from 2006, it increased steadily until 2009, where there was a sudden drop of publications for that year. After 2009, there was a big increase, reaching the peak in 2010 \& 2011. There was again a sudden drop in 2015, with increases from 2016 to 2019. Since the paper list was gathered in early 2020, there may be papers published after our search. Overall, there is a considerable increase in the number of studies from 2010, though fluctuating over this time.

\begin{table}[]
\centering
\caption{List of relevant studies for answering each RQ}
\label{TABLE 9: Studies for RQ}
\resizebox{\columnwidth}{!}{%
\begin{tabular}{@{}clllc@{}}
\toprule
\textbf{RQ} & \multicolumn{3}{c}{\textbf{Paper ID}}                                                                                                                                                                                                                                                                                                                                                                                                                                                                                                                                            & \textbf{Paper count} \\ \midrule
RQ1         & \multicolumn{3}{l}{\begin{tabular}[c]{@{}l@{}}IEE01, IEE02, IEE03, IEE04, IEE05, IEE06, IEE07, IEE08, IEE09, \\ IEE10, IEE11, IEE12, IEE13, IEE14, IEE15, IEE16, IEE17, IEE18, \\ IEE19, IEE20, IEE21, IEE22, IEE23, IEE24, IEE25, IEE26, IEE27, \\ ACM01, ACM02, ACM03, ACM04, ACM05, ACM06, ACM07, \\ ACM08, ACM09, ACM10, ACM11, ACM12, ACM13, ACM14, \\ ACM15, ACM16, ACM18, ACM19, ACM20, SP01, SP02, SP03, \\ SP04, SP05, SP06, SP07, SP08, SP09, SP10, SP11, SP12, SP13, \\ SP15, SP16, SP17, WI01, WI02, WI04, SB01, SB02, SB03, SB04, \\ SB05, SB06, SB07, SB08 \end{tabular}} & 74                   \\ \\
RQ2         & \multicolumn{3}{l}{\begin{tabular}[c]{@{}l@{}}IEE01, IEE02, IEE03, IEE04, IEE06, IEE07, IEE09, IEE10, IEE11, \\ IEE12, IEE13, IEE14, IEE15, IEE16, IEE17, IEE18, IEE19, IEE20, \\ IEE22, IEE23, IEE24, IEE25, IEE26, IEE27, ACM01, ACM02, \\ ACM04, ACM05, ACM06, ACM11, ACM15, ACM18, ACM19, \\ ACM20, SP01, SP02, SP03, SP04, SP05, SP06, SP07, SP08, SP09, \\ SP12, SP13, SP16, SP17, WI04, SB01, SB02, SB05, SB7, SB08\end{tabular}}                                                                                                                                               & 53                   \\ \\
RQ3         & \multicolumn{3}{l}{\begin{tabular}[c]{@{}l@{}}IEE01, IEE02, IEE10, IEE11, IEE12, IEE13, IEE17, IEE21, ACM01, \\ ACM03, ACM19, ACM20, SP02, SP03, SP04, SP16, WI01, SB01, \\ SB03, SB04, SB05, SB06, SB07\end{tabular}}                                                                                                                                                                                                                                                                                                                                                           & 23                   \\ \\
RQ4         & \multicolumn{3}{l}{\begin{tabular}[c]{@{}l@{}}IEE01, IEE02, IEE03, IEE04, IEE06, IEE07, IEE08, IEE09, IEE10, \\ IEE11, IEE12, IEE13, IEE14, IEE16, IEE17, IEE18, IEE19, IEE20, \\ IEE22, IEE23, IEE24, IEE25, IEE26, IEE27, ACM02, ACM04, \\ ACM11, ACM15, ACM18, SP01, SP02, SP03, SP04, SP07, SP12, \\ WI04, SB02, SB04, SB05, SB07, SB08\end{tabular}}                                                                                                                                                                                                                              & 41                   \\ \bottomrule
\end{tabular}}
\end{table}

The questions in our google form used for the primary study paper data extraction were prepared based on RQs. Table \ref{TABLE 9: Studies for RQ} shows the list of primary studies which we found relevant for answering each RQ. In the followings subsections we discuss answers to each of our RQs in turn.

\subsection{What is the motivation behind each primary study on identifying the effects of human aspects in requirements engineering (RQ1)?}

\begin{itemize}

    \item \textit{The goals/objectives/motivation behind each study reviewed}
\end{itemize}

\par Each research paper reviewed in our SLR has its' own goals, objectives and motivation. We have categorized these for each study according to three main aspects. This is based on how the studies claimed to be planning to address the problem and how the research was going to be done. 

\textbf{I. To identify/investigate the effects of human aspects on RE:}  Since this is the main reason for most of the studies to conduct their research, most of the human aspects from the categories shown in Figure \ref{Human_Factors} are concerned with identifying their effect on the RE process. Investigating the effects can be specified as, identifying the challenges in RE process due to human aspects (eg: [IEE02], [IEE17], [ACM07], [ACM09], [SP01]); better understanding of the support/influence of human aspects and different viewpoints of it (eg: [IEE07], [IEE11], [IEE14], [ACM13], [SP04], [SP09], [WI01]); finding evidence/key contribution of different human aspects to the RE process (eg: [IEE15], [ACM02], [ACM03], [ACM15], [SP03]); or to analyse overall impact of human aspects on the RE process (eg: [IEE20], [IEE22], [IEE25], [ACM01], [ACM05], [SB01]) 

\textbf{II. Present a novel model/framework/approach to improve RE considering human aspects:} This is the second most common reason to conduct the studies. In this category, the studies focus on presenting novel approaches to improve the RE process by better considering or  incorporating human aspects. The approaches include introducing a systematic process or a technique (eg: [IEE01], [IEE05], [ACM04], [SP06], [SP11]); defining set of strategies ([SP10], [SB02]); developing a theoretical or working model (eg: [IEE12], [ACM20], [SP07], [SP08] ,[WI04], [SB05],[SB07]); or producing a tool (eg: [IEE03], [ACM13])  that will eventually assist in analysing human aspects for improving the quality of the RE process.    

\textbf{III. Incorporate existing models/tools in RE considering human aspects:} Considerably less studies have considered incorporating an existing model or a tool into the RE process as a goal/motivation to conduct their study. Among the papers that have considered incorporating an existing model or a tool, the main focus is usually to identify current models, tools, techniques or approaches in other domains and check whether those can be adapted to identify and analyse the effect of human aspects in the RE process. Hence, the adaptability of both technical approaches and psychological models have been examined in these papers. In [IEE06], a Spatial Hypertext Wiki was investigated as a collaborative tool for supporting creativity in the RE process. [SB04] used a personality testing method (Myers-Bridge Type Indicator (MBTI)) to uncover the different personality types in a software team. 

\begin{itemize}
    \item \textit{The target domain of the studies: academic or industry.}
\end{itemize}

\begin{figure}
\centering
  \includegraphics[width=\linewidth]{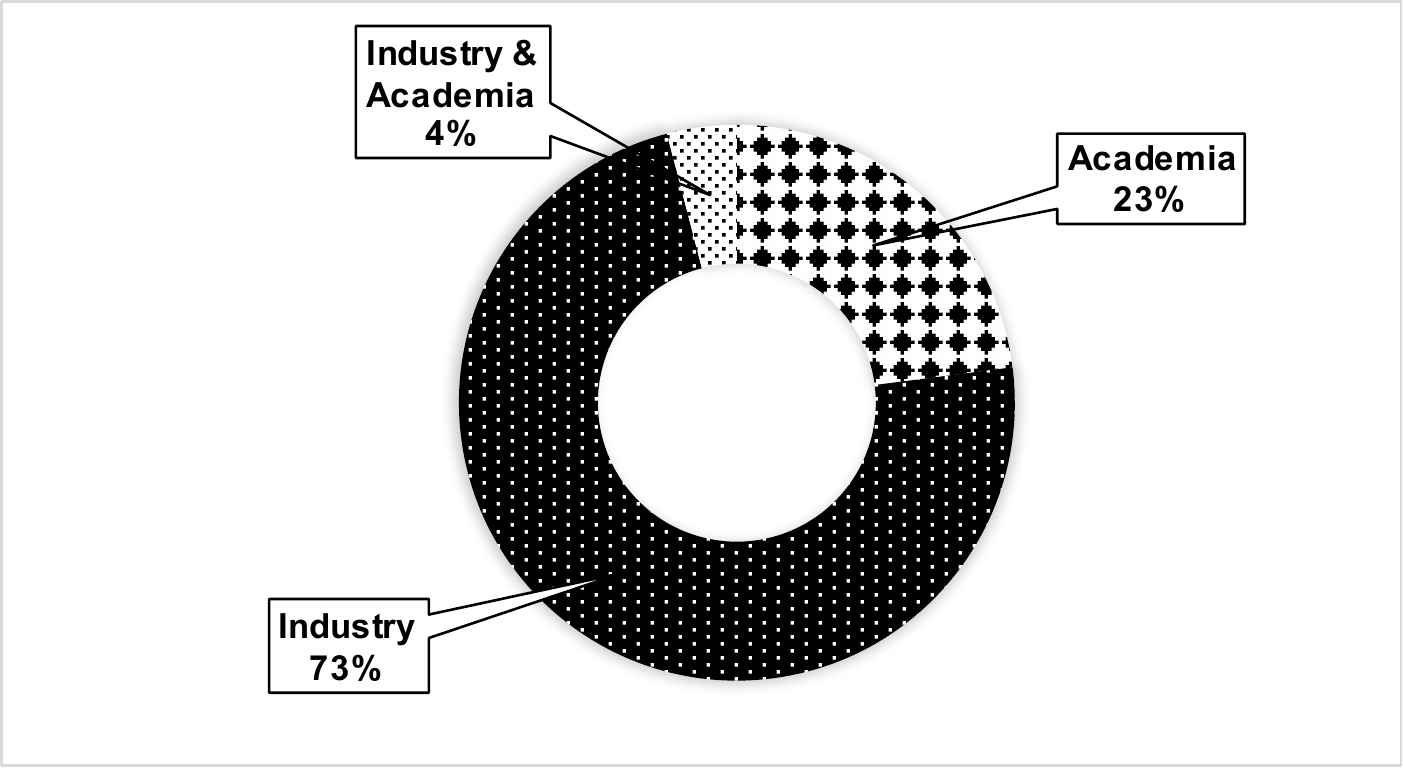}
  \caption{Target areas of the study}
  \label{RE_Context}
\end{figure}

\begin{figure}
  \includegraphics[width=\linewidth]{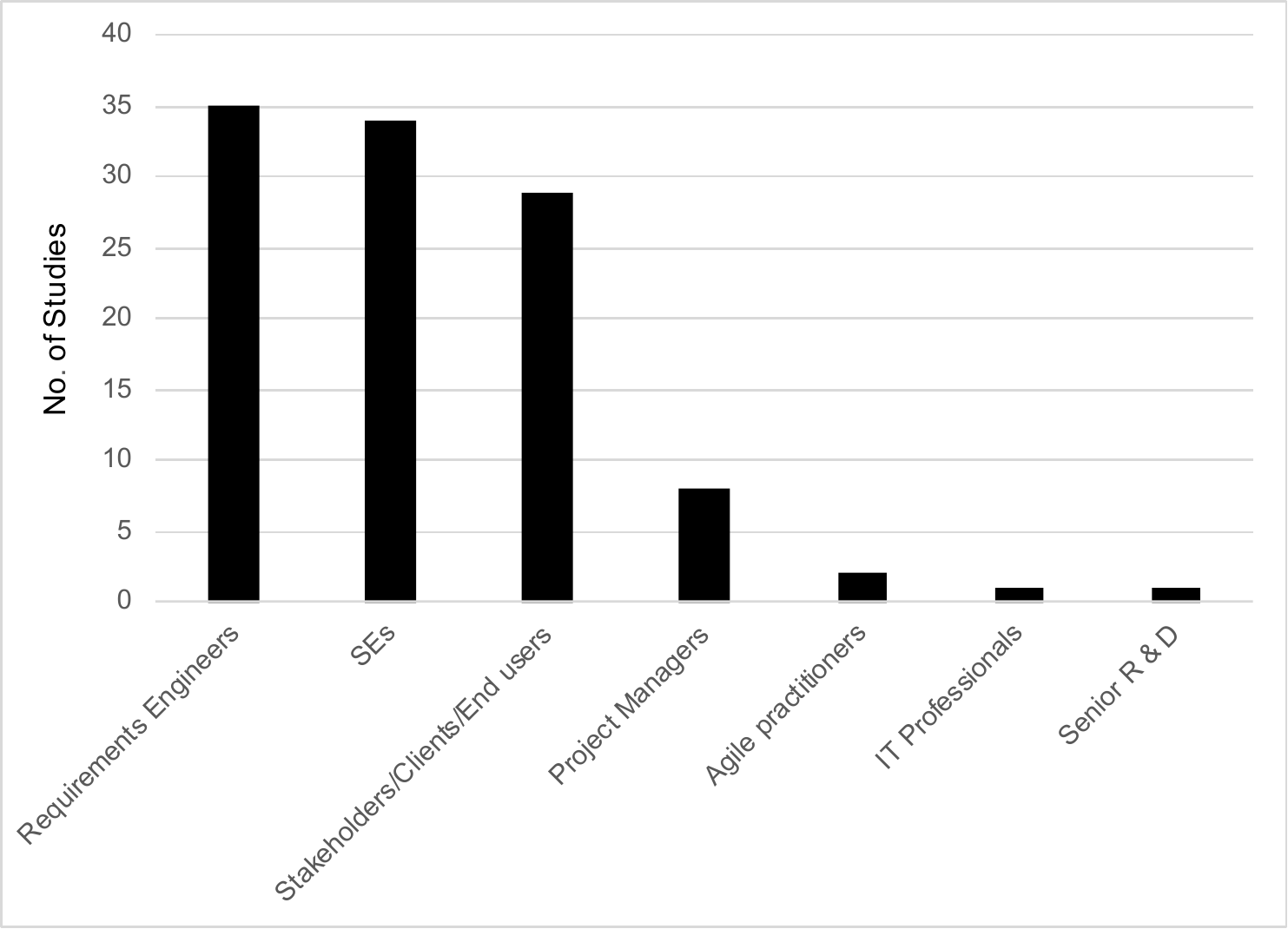}
  \caption{Subjects used in studies}
  \label{SubjectsofStudy}
\end{figure}

Shown in Figure \ref{RE_Context}, the subject of most studies are related to the software industry with 73\%,  compared to 23\% of studies focused only on academia, and only 4\% of studies considering both a software industry and academia domain. Figure \ref{SubjectsofStudy} shows the variety of the subjects considered within the software industry focused studies. The highest number of studies (35) have considered only the requirements engineering process and requirements engineers, whereas the second highest number of studies (34) have investigated more general SE processes, including RE as one phase. The rest focus on stakeholder issues (29 studies), i.e. those  who are external to software development team -- project managers (8 studies), agile practitioners (2 studies), IT professionals (1 study) and senior R \& D personnel (1 study). In these papers they have investigated requirements engineering only as a part of the overall study.\\


\begin{table}[]
\centering
\caption{Studies based on one or two human aspects}
\label{TABLE 12: Focused study areas}
\resizebox{\columnwidth}{!}{%
\begin{tabular}{@{}ll@{}}
\toprule
\multicolumn{1}{c}{\textbf{Human Aspect}}     & \multicolumn{1}{c}{\textbf{Paper ID}}      \\ \midrule
Communication issues                          & ACM07, ACM11, SP01, SP03, SP05, SP16, SB02 \\
Personality                                   & IEE25, IEE05, ACM02, ACM04, ACM15, SB04    \\
Human values                                  & IEE06, IEE07, IEE23, ACM13, SP04           \\
Gender                                        & IEE18, ACM10, ACM16                        \\
Motivation                                    & ACM05, ACM06, WI01                         \\
Emotions \& motivation                        & IEE10, IEE12, ACM08                        \\
Emotions                                      & WI03, ACM19                                \\
Personality \& Communication issues           & SP06, SP07                                 \\
Culture                                       & IEE24                                      \\
Domain knowledge                              & IEE14                                      \\
Human errors                                  & IEE16                                      \\
Physical issues                               & SP08                                       \\
Community values                              & ACM09                                      \\
Team maturity                                 & ACM12                                      \\
Task workload                                 & IEE03                                      \\
Personality \& Human values                   & IEE25                                      \\
Personality \& Motivation                     & IEE26                                      \\
Personality \& Emotions                       & SB01                                       \\
Personality \& Culture                        & ACM20                                      \\
Personality \& Attitude                       & SB03                                       \\
Personality \& Selfmanagement                 & SP17                                       \\
Culture \& Communication issues               & IEE20                                      \\
Culture \& Motivation                         & ACM03                                      \\
Culture \& Gender                             & IEE11                                      \\
Culture \& Geographic distribution/ Time zone & IEE08                                      \\
Emotions \& Gender                            & ACM18                                      \\
Communication issues \& Domain knowledge      & WI02                                       \\
Communication issues \& Emotions              & SP02                                       \\
Human values \& Knowledge sharing             & SP09                                       \\ \bottomrule
\end{tabular}%
}
\end{table}

\cornersize{.2} 
\ovalbox{\begin{minipage}{7.5cm}
\small
\textbf{Answers to RQ1:} To date, the aims of most studies of human aspects impacting RE have been focusing on investigating which human aspects impact RE. Considerably less studies have focused on new models and approaches to improve RE based on these human aspects, and very few have focused on incorporating existing models or tools into RE. Most studies have however been done with industry-based projects.
\end{minipage}}

\subsection{What is the current status of the research studies on the effect of human aspects on requirements engineering (RQ2)?} \label{4.3.2}

\begin{itemize}
    \item \textit{Human Aspect impacts on RE investigated to date}
\end{itemize}

This research question focuses on identifying and categorizing the human aspects that have been investigated to date. From our data extraction and analysis of 74 primary studies, we see that these studies have focused on a range of human aspects. The majority of the studies focused on one human aspect (33 studies), and others considered two, three or up to a maximum of four different human aspects. 

For each human aspect, we have considered its definition and grouped similar aspects together. For this we have used definitions that have generally been considered in the SE context and those in the selected primary studies. The final set of human aspects we identified were categorized in to three groups -- \textbf{individual} related human aspects, \textbf{technical} related human aspects and \textbf{team} related human aspects with the purpose of categorizing the broadly different human aspects studied to date. Furthermore, we used the "technical" related human aspects category combining with individual and team related human aspects to categorize human aspects that we felt are not fully appropriate to individual or team categories. These are summarised in Figure \ref{Human_Factors}. Among our selected primary studies, the highest number of studies were about individual- and team-related human aspects. Only 7 studies were about technical-related human aspects.
As shown in the Table \ref{TABLE 12: Focused study areas}, 52 out of 74 primary studies have only focused on one or two human aspects. The other 22 studies discussed more than two aspects in the studies. 
\begin{figure}
  \includegraphics[width=\linewidth]{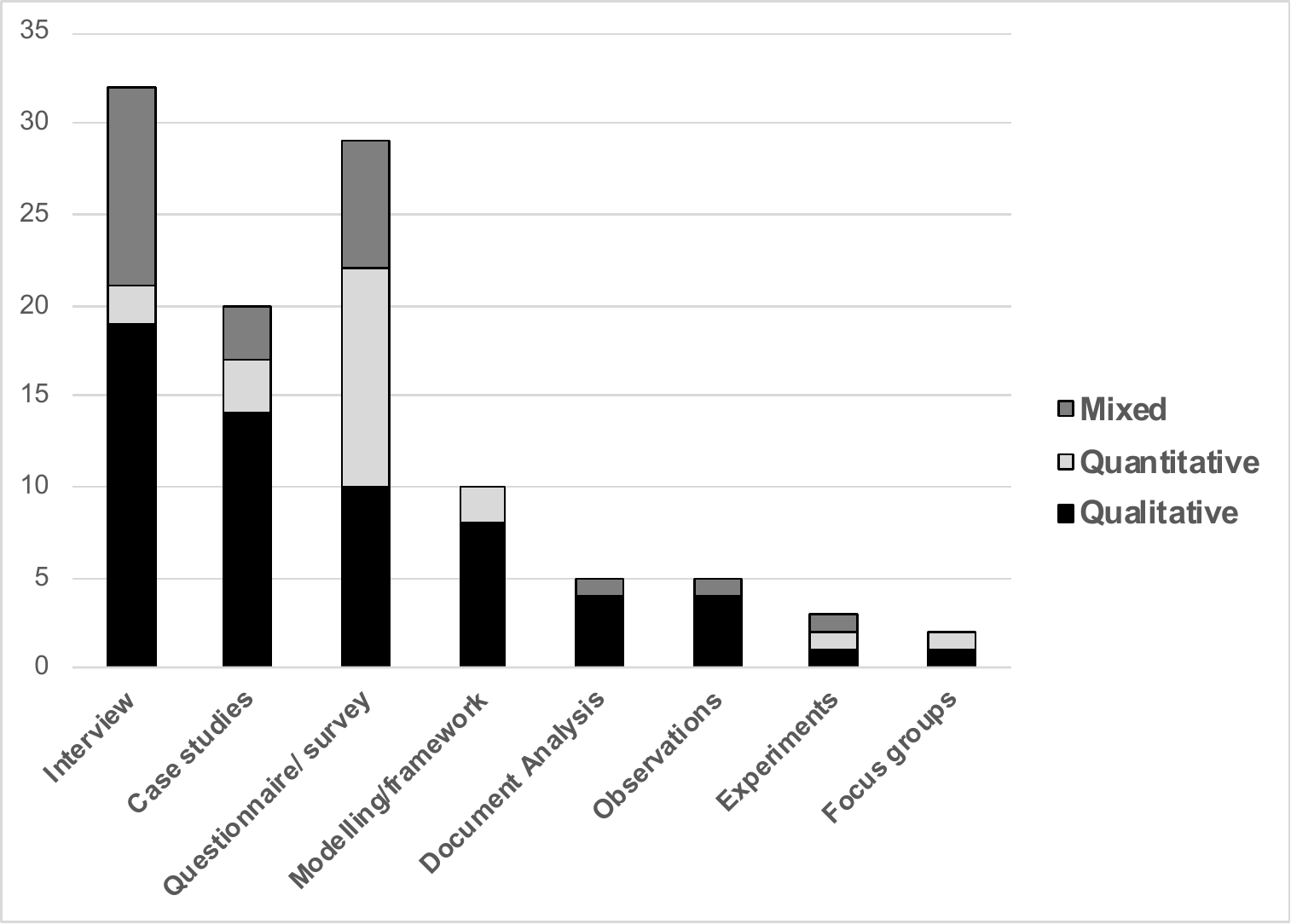}
  \caption{Research method over type}
  \label{Method}
\end{figure}
\begin{figure*}[t!]
 \centering
  \includegraphics[width=0.75\linewidth]{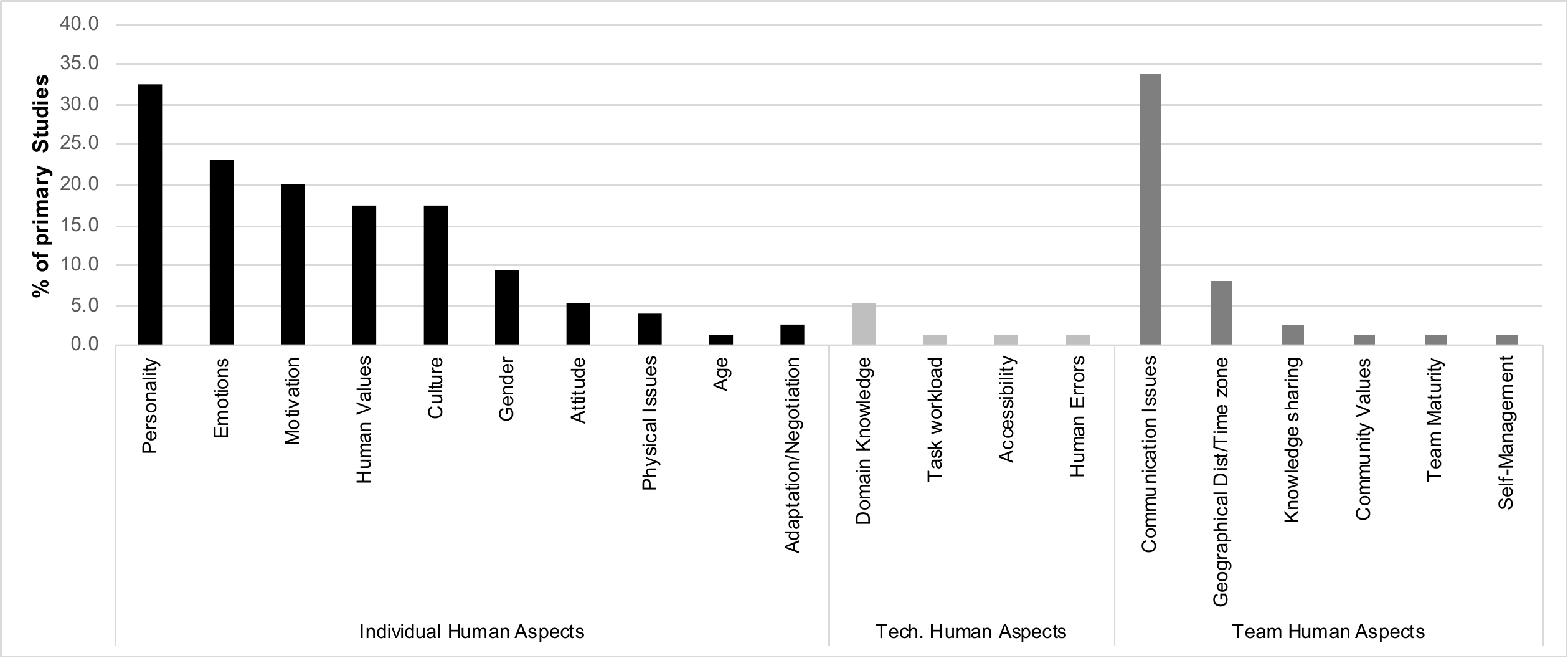}
  \caption{Categorization of the human aspects/human-centric (HC) issues studied}
  \label{Human_Factors}
\end{figure*}


Considering the set of individual-related human aspects, 32.4\% of studies were about \emph{personality} and 23\% of studies were about \emph{emotions}. 20.3\%, 17.6\% and 17.6\% were about \emph{motivation, human values} and \emph{culture} respectively. According to Schwartz's definition of human values \cite{RN2687} \cite{RN2965}, creativity is considered as a human value and so we have grouped the 10 studies that have focused on creativity under ``human values". Therefore, this 17.6\% includes studies of \emph{human values} including creativity. The rest of studies focused on other individual human aspects like \emph{gender} (9.5\%), \emph{attitude} (5.4\%), \emph{physical issues} (4.1\%), \emph{adaptation/negotiation} (2.7\%). and \emph{age} (1.4\%)  

The second highest number of studies were about team related human aspects.  33.8\% of the studies were about \emph{communication issues} .
Other aspects considered were \emph{geographical distribution} (8.1\%), \emph{knowledge sharing} (2.7\%), \emph{community values} (1.4\%), \emph{team maturity} (1.4\%) and \emph{self-management} (1.4\%) were considered in far fewer of the selected primary studies. We identified that there were very few studies on technical related human aspects. Among these the majority were about \emph{domain knowledge} (5.4\%), and the rest were about \emph{task workload} (1.4\%), \emph{accessibility} (1.4\%) and \emph{human errors} (1.4\%).

\begin{itemize}
    \item \textit{Methodologies used to identify the effect of human aspects on RE}
\end{itemize}

Authors of the selected primary studies have used a variety of research techniques, some of them multiple ones. Among the described techniques, \emph{interview} (43.24\%), \emph{questionnaire/surveys} (39.19\%), \emph{case studies} (27.03\%) and \emph{modelling} (13.51\%) were the most used research techniques. \emph{Document analysis} (6.76\%), \emph{observations} (6.76\%), \emph{experiments} (4\%) and \emph{focus groups} (2.70\%) were less common.
Figure \ref{Method} summarises the research techniques used in the reviewed papers. We can see that the majority of the studies (82.43\%) have used \emph{qualitative techniques} whereas 28.38\%  have conducted the studies \emph{quantitatively}, and 32\%  of studies have used a mixed approach. 


\begin{table}[]
\centering
\caption{Existing domain models used in primary studies}
\label{TABLE 10: Domain Models}
\resizebox{\columnwidth}{!}{%
\begin{tabular}{@{}cll@{}}
\toprule
\footnotesize
\textbf{Human Aspect}          & \multicolumn{1}{c}{\textbf{Domain Models}}                                         & \multicolumn{1}{c}{\textbf{Paper ID}}                                \\ \midrule
\multirow{4}{*}{Personality}   & Five Factor Model (FFM)                                                            & \begin{tabular}[c]{@{}l@{}}SP11, SP17, \\ SB01, SB06\end{tabular}    \\
                               & Myers-Bridge Type Indicator (MBTI)                                                 & \begin{tabular}[c]{@{}l@{}}ACM02, ACM04, \\ ACM15, SB04\end{tabular} \\
                               & International Personality Item Pool (IPIP)                                         & IEE25, SB03                                                          \\
                               & HEXACO personality Inventory                                                       & IEE04                                                                \\ \midrule
\multirow{2}{*}{Culture}       & Hofstede Model                                                                     & \begin{tabular}[c]{@{}l@{}}IEE11, IEE15, \\ IEE24\end{tabular}       \\
                               & Trauth’s theoretical framework                                                     & ACM01                                                                \\ \midrule
\multirow{4}{*}{Emotions}      & Clore \& Collins (OCC) model                                                       & IEE12, IEE13                                                         \\
                               & Self-Assessment manikin (SAM)                                                      & ACM08                                                                \\
                               & Satisfaction. Gladstein’s questionnaire                                            & SP01                                                                 \\
                               & Trait Emotional Intelligence Questionnaire                                         & SB06                                                                 \\ \midrule
\multirow{6}{*}{Motivation}    & Maslow’s motivation theory                                                         & IEE12, IEE13, IEE26                                                  \\
                               & Motivation-Hygiene Theory                                                          & ACM05                                                                \\
                               & Motivation-Skinner’s behaviorist theory                                            & IEE26                                                                \\
                               & The MOCC model                                                                     & ACM06                                                                \\
                               & \begin{tabular}[c]{@{}l@{}}An Intrinsic Motivation \\ Inventory (IMI)\end{tabular} & ACM08                                                                \\
                               & Expectancy Theory                                                                  & ACM05                                                                \\ \midrule
\multirow{3}{*}{Communication} & \begin{tabular}[c]{@{}l@{}}Influence on Consensus \\ (IC) Model\end{tabular}       & SP02                                                                 \\
                               & \begin{tabular}[c]{@{}l@{}}Online Communication \\ Model (OCM)\end{tabular}        & SP17                                                                 \\
                               & \begin{tabular}[c]{@{}l@{}}Typology of Communication \\ challenges\end{tabular}    & SP01                                                                 \\ \midrule
\multirow{3}{*}{Human values}  & \begin{tabular}[c]{@{}l@{}}Schwartz’s universal values \\ model\end{tabular}       & ACM13                                                                \\
                               & CPS Method (creativity)                                                            & \begin{tabular}[c]{@{}l@{}}IEE07, \\ IEE25\end{tabular}              \\
                               & \begin{tabular}[c]{@{}l@{}}Componential Model of \\ Creativity (CMoC)\end{tabular} & IEE21                                                                \\ \midrule
\multirow{2}{*}{Human Errors}  & Human Error Taxonomy (HET)                                                         & IEE16                                                                \\
                               & \begin{tabular}[c]{@{}l@{}}Human Error Abstraction \\ Assist (HEAA)\end{tabular}   & IEE16                                                                \\
Team Maturity                  & \begin{tabular}[c]{@{}l@{}}Team Evolution and \\ Maturation (TEAM)\end{tabular}    & ACM12                                                                \\ \bottomrule
\end{tabular}%
}
\end{table}

\begin{itemize}
    \item \textit{Types of existing domain models used to identify the effect of human aspects on RE}
\end{itemize} 
Several studies have used existing psychology domain models to identify the effect of human aspects on RE. Most of these models have originated in the psychology research domain quite some time ago, but more recently used in the SE research domain. Among our 74 primary studies, 51.4\% have considered existing domain models in their studies, whereas 48.7\% of studies do not seem to have used any existing domain models related to human aspects in their studies. 

Table \ref{TABLE 10: Domain Models} summarises our analysis. Several models have been used to identify the effects of same human aspects in several different studies. To identify the effects of \textbf{personality}, 4 studies have used \emph{Five Factor Model (FFM)} and another 4 studies have used \emph{Myers-Bridge Type Indicator (MBTI)} model. 2 studies have used \emph{IPIP (International Personality Item Pool)}, a short measure of FFM. 

For \textbf{culture}, 3 studies used the \emph{Hofstede model}, and 1 \emph{Trauth’s theoretical framework}. For \textbf{emotions}, 4 types of domain models have been used. \emph{Clore \& Collins (OCC)} covers a broad range of emotions (22 emotions) and is used by 2 studies. With regards to \textbf{motivation}, 6 types of motivation theories have been drawn up in the primary studies. Among these \emph{Maslow’s motivation theory} is the most used one by 3 studies.

Regarding the \textbf{communication}, 3 types of domain models have been used in 3 separate studies. For \textbf{Human values}, \emph{Schwartz’s universal values model}, the \emph{Creativity Problem Solving (CPS)} method, and \emph{Componential Model of Creativity (CMoC)}. The models used for \textbf{human errors} are \emph{Human Error Taxonomy (HET)} and \emph{Human Error Abstraction Assist (HEAA) \& training procedures}, proposed by  J. Reason\cite{RN2963} and V. K. Anu\cite{RN2964}. These two models have been used by the same primary study [IEE16]. 

The \emph{Team Evaluation and Maturation (TEAM) model} proposed by Morgan, Salas e Glickman \cite{RN2962} has been used to identify the effects of \textbf{team maturity}, and is used widely in several disciplinary areas including psychology and SE. Among the primary studies [IEE12], [IEE13], [IEE25], [IEE26], [ACM05], [SP01] and [SP17] have all used more than one domain model in the same study, as the studies have focused more than one human aspect.

\begin{itemize}
    \item \textit{Presented solutions to address the effect of human aspects on the  RE process}
\end{itemize}

Almost all of the 74 primary studies propose some form of solution related to the effect of particular human aspect(s) considered in the study. We have categorized these solutions into five main areas considering the solutions presented. A proposed solution can be a working model/tool, a theoretical model/framework/prototype, an approach/strategy/guideline to address human aspects in RE, identification of relationships among human aspects in RE, or revealing new challenges/effects of human aspects on RE. 

 \renewcommand\labelitemii{$\square$}
\begin{itemize}
    \item[$\square$] \textbf{Solution as a working model/tool or a theoretical model/prototype}
\end{itemize}

\par Among the solutions presented in our primary studies, 44.5\% propose a new or modified framework/model, and this includes both working model or a theoretical model. A summary of the range of solutions presented in the  primary studies can be seen in Table \ref{tab:solutions}. Considering working models, we describe these as a totally new model, extended version of an existing model, or a tool that  incorporates and implements an existing model with the aim of identifying or supporting the effect of human aspect(s) in RE. For example, in [IEE01] a psychological-driven goal model has been presented targeting on \emph{human values, motivation and emotions} aspects aiming at improving RE by capturing more requirements that cannot be elicited using traditional RE techniques. RAMSET: a Role Assignment Methodology for SE Teams is a working model presented in [ACM04]. It is based on \emph{personality} and used socio-metric techniques and psycho-metrics to support team assignment, including for RE tasks. In [IEE23], a theoretical model has been suggested for requirements engineers incorporating their personality and communication which will assist to select appropriate elicitation technique in a cooperative distributed environment. [IEE13] is another study that presented a theoretical framework (Taxonomy) to deal with soft issues in RE, including \emph{personality, emotions, motivation and attitude}. The authors claim this can be used in two modes to suit novice or expert analysts.   

\begin{table*}[]
\caption{Solutions presented to address the effect of human aspects on RE}
\label{tab:solutions}
\resizebox{\textwidth}{!}{%
\begin{tabular}{@{}lllll@{}}
\toprule
\textbf{Solution Type}                                                                   & \textbf{Solution}                                                                                                                                                       & \textbf{Target Human Aspect}                                                                                        & \textbf{Impact on RE}                                                                                                                                                                                                                                                                                                                                                                                   & \textbf{Paper ID} \\ \midrule
\multirow{15}{*}{Working model/Tool}                                                      & A psychological-driven goal model                                                                                                                                       & Human values, Motivation, Emotions                                                                                  & \begin{tabular}[c]{@{}l@{}}Use to improve RE by capturing more requirements that cannot \\ be elicited using traditional RE techniques\end{tabular}                                                                                                                                                                                                                                                     & IEE01             \\
                                                                                         & Value Based RE (VBRE)                                                                                                                                                   & Human values, Motivation, Emotions                                                                                  & Systematically focuses on socio-political issues in RE                                                                                                                                                                                                                                                                                                                                                  & SB05              \\
                                                                                         & Statistical model implemented using R                                                                                                                                   & Emotions                                                                                                            & \begin{tabular}[c]{@{}l@{}}Helps to identify affects (emotions) linked to cognitive processing \\ activities and individual productivity\end{tabular}                                                                                                                                                                                                                                                  & WI03              \\
                                                                                         & People-oriented software engineering (POSE) model                                                                                                                       & Emotions                                                                                                            & \begin{tabular}[c]{@{}l@{}}Capability to carry the voice of the user and keep many \\ stakeholder groups involved throughout RE\end{tabular}                                                                                                                                                                                                                                                   & ACM19             \\
                                                                                         & A three layer model                                                                                                                                                     & \begin{tabular}[c]{@{}l@{}}Communication, Culture, Knowledge \\ management, \\ Geographic distribution\end{tabular} & \begin{tabular}[c]{@{}l@{}}Describes the four major challenges related to target human aspects;\\  inadequate communicaction, cultural diversity, ineffective knowledge \\ management and time differences, shows the difficulties created by \\ these challenges and identifies impact of these challenges in RE\end{tabular}                                                                            & IEE02             \\
                                                                                         & REIM- A training method for requirements engineers                                                                                                                      & \begin{tabular}[c]{@{}l@{}}Personality, Emotions, Communication, \\ Adaptation/negotiation\end{tabular}             & \begin{tabular}[c]{@{}l@{}}Train requirements engineers to detect underlying soft issues \\ (target human aspects) that influence or even hinder factual \\ clarification in RE\end{tabular}                                                                                                                                                                                                            & IEE19             \\
                                                                                         & A competency model with 16 critical competencies                                                                                                                        & \begin{tabular}[c]{@{}l@{}}Personality, Emotions, Motivation, \\ Communication\end{tabular}                         & \begin{tabular}[c]{@{}l@{}}16 Critical competencies integrate contextual and situational factors that\\  are important for the of requirements analyst\end{tabular}                                                                                                                                                                                                                                 & WI04              \\
                                                                                         & \begin{tabular}[c]{@{}l@{}}HEAA (Human Error Abstraction Assist) -training \\ procedure and intervention tool\end{tabular}                                              & Human Errors                                                                                                        & \begin{tabular}[c]{@{}l@{}}Used to train industry software practitioners about the human errors that\\  frequently occur during the RE process\end{tabular}                                                                                                                                                                                                                                              & IEE16             \\
                                                                                         & A blind user RE method                                                                                                                                                  & Pysical issues                                                                                                      & Use to develop the requirements for mobile services tailored to blind users                                                                                                                                                                                                                                                                                                                             & SPO8              \\
                                                                                         & \begin{tabular}[c]{@{}l@{}}HUCRE method and two support tools (workload \\ analyser  \& functional allocation advisor)\end{tabular}                                     & Task workload                                                                                                       & \begin{tabular}[c]{@{}l@{}}Contribute to the RE process by analysing human-related non-functional\\ requirements such as workload, reliability and decision-making effectiveness\end{tabular}                                                                                                                                                                                                           & IEE03             \\
                                                                                         & \begin{tabular}[c]{@{}l@{}}RAMSET; A Role Assignment Methodology for \\ Software Engineering\end{tabular}                                                               & Personality                                                                                                         & \begin{tabular}[c]{@{}l@{}}Support team assignment including RE tasks considering socio-metric \\ techniques and psycho-metrics techniques\end{tabular}                                                                                                                                                                                                                                                 & ACM04             \\
                                                                                         & Behavioral Simulator                                                                                                                                                    & Personality, Communication                                                                                          & \begin{tabular}[c]{@{}l@{}}Train soft skills which are personality-driven abilities related to the \\ emotive and communicative sphere\end{tabular}                                                                                                                                                                                                                                                     & SP07              \\
                                                                                         & ShyWiki Tool; Spatial Hypertext Wiki                                                                                                                                     & Creativity (Human values)                                                                                           & Assist as a collaborative tool for supporting creativity in the RE process                                                                                                                                                                                                                                                                                                                              & IEE06             \\
                                                                                         & Requirements elicitation issues model                                                                                                                                   & Communication                                                                                                       & \begin{tabular}[c]{@{}l@{}}Provides an empirical perspective on the impacts of elicitation issues along with \\ priority-setting of elicitation issues. The priority setting of  parameters  can  \\ support  business  analyst/requirements engineers   to   be   more   prepared   to   \\ realize   and  address relevant  risks  that  may  potentially  surface  during  elicitation.\end{tabular} & SP05              \\
                                                                                         & A web-based competency evaluation platform                                                                                                                              & Emotions, Motivation                                                                                                & \begin{tabular}[c]{@{}l@{}}Needs to be supplemented with qualitative data and assist gathering information about \\ motivation emotions of the employees when involving in the tasks\end{tabular}                                                                                                                                                                                                        & ACM08             \\ \midrule
\multirow{14}{*}{\begin{tabular}[c]{@{}l@{}}Theoretical model/\\ Prototype\end{tabular}} & \begin{tabular}[c]{@{}l@{}}Conceptual mapping model Based on MBTI \\ personality model\end{tabular}                                                                     & Personality                                                                                                         & \begin{tabular}[c]{@{}l@{}}Support to identify personality of software practitioners (including RE) \\ via conceptual mapping - assist towards role assignment\end{tabular}                                                                                                                                                                                                                             & ACM02             \\
                                                                                         & A model that distinct personality patterns                                                                                                                              & Personality                                                                                                         & \begin{tabular}[c]{@{}l@{}}Qualitatively identifies the technology affordance analysing larger set of people with \\ their distinct personality patterns\end{tabular}                                                                                                                                                                                                                                   & IEE05             \\
                                                                                         & A model for requirements engineers                                                                                                                                      & Personality, Communication                                                                                          & \begin{tabular}[c]{@{}l@{}}Assist to select an elicitation technique in a cooperative distributed environment \\ based on their language, priorities and values\end{tabular}                                                                                                                                                                                                                             & SP06              \\ 
                                                                                         & RE framework for service engineering                                                                                                                                    & Creativity (Human values)                                                                                           & \begin{tabular}[c]{@{}l@{}}Service RE process in two phases - requirements specification and requirements \\ evolution phases in service-oriented RE\end{tabular}                                                                                                                                                                                                                                       & IEE23             \\
                                                                                         & An approach for value study                                                                                                                                             & Human values                                                                                                        & \begin{tabular}[c]{@{}l@{}}Consider values as mental representations which influence technical outcomes, \\ tensions and relationships in SE including RE\end{tabular}                                                                                                                                                                                                                                  & ACM13             \\
                                                                                         & Conceptual model related to creativity                                                                                                                                  & Creativity (Human values)                                                                                           & \begin{tabular}[c]{@{}l@{}}Focus on three contextual factors explain why creativity can be understood \\ differently in RE and five dimensions that explain qualitatively how creativity's \\ meaning can vary in RE\end{tabular}                                                                                                                                                                       & SP09              \\
                                                                                         & RepGrid (Repertory Grid Technique)                                                                                                                                      & Communication                                                                                                       & \begin{tabular}[c]{@{}l@{}}Suggests to incorporate repertory grid technique with other elicitation tools to \\ identify key communication issues in elicitation\end{tabular}                                                                                                                                                                                                                            & SB02              \\
                                                                                         & \begin{tabular}[c]{@{}l@{}}Effective Situation Requirements Template - \\ early prototype\end{tabular}                                                                  & Emotions, Motivation                                                                                                & \begin{tabular}[c]{@{}l@{}}Provides a support for situations/problems in RE related to emotions and \\ motivation\end{tabular}                                                                                                                                                                                                                                                                          & IEE12             \\
                                                                                         & Analysis method to deal with soft issues in RE                                                                                                                          & \begin{tabular}[c]{@{}l@{}}Personality, Emotions, Motivation, \\ Attitude\end{tabular}                              & \begin{tabular}[c]{@{}l@{}}Introduce new considerations into the RE process by drawing attention to values, \\ motivations and emotion - can be used in novice and expert analysis\end{tabular}                                                                                                                                                                                                         & IEE13             \\
                                                                                         & \begin{tabular}[c]{@{}l@{}}A DRASIS (Dynamic Role Allocation Support \\ In Software engineering)\end{tabular}                                                           & \begin{tabular}[c]{@{}l@{}}Personality, Culture, Geographic \\ Distribution\end{tabular}                            & \begin{tabular}[c]{@{}l@{}}Develop for dynamic role allocation in software engineering groups -  Identify the \\ effects that cultural differences and personality characteristics have on dynamic \\ role allocation\end{tabular}                                                                                                                                                                      & IEE15             \\
                                                                                         & \begin{tabular}[c]{@{}l@{}}A framework based on questionnaire design to measure \\ the software engineers (including requirements \\ engineers) motivation\end{tabular} & Motivation                                                                                                          & \begin{tabular}[c]{@{}l@{}}May serve as a generic tool for measurement of motivation, which can be easily \\ reused in future research or in practice\end{tabular}                                                                                                                                                                                                                                      & ACM05             \\
                                                                                         & A framework related to job satisfaction/ motivation                                                                                                                     & Motivation/ Job satisfaction                                                                                        & \begin{tabular}[c]{@{}l@{}}May assist to identify job satisfaction/ motivation  of software engineers referring \\ to distinct phenomena\end{tabular}                                                                                                                                                                                                                                                   & ACM06             \\
                                                                                         & Conceptual framework base on the literature                                                                                                                             & Culture, Communication                                                                                              & \begin{tabular}[c]{@{}l@{}}Identify the effect of  communication in requirements elicitation process, Trust, \\ Interpersonal skills, Organizational culture, knowledge during elicitation\end{tabular}                                                                                                                                                                                                 & SP02              \\
                                                                                         & \begin{tabular}[c]{@{}l@{}}A framework for requirements elicitation process \\ in global software projects\end{tabular}                                                 & \begin{tabular}[c]{@{}l@{}}Culture, Communication, \\ Geographic Distribution\end{tabular}                          & \begin{tabular}[c]{@{}l@{}}Focusing on problem prediction and different strategies to avoid or decrease \\ their impact on GSD project performance\end{tabular}                                                                                                                                                                                                                                         & SB07              \\ \cmidrule(l){1-5} 
\end{tabular}%
}
\end{table*}

 \renewcommand\labelitemii{$\square$}
\begin{itemize}
    \item[$\square$] \textbf{An approach/strategy/guideline to address human aspects in RE}
\end{itemize}

Among the 55.4\% of studies that do not present a working or a theoretical model as a solution, many have presented an approach, set of strategies or guidelines might be used to address the effect of human aspects on RE. [IEE07] has presented such approach which is about mapping of \emph{creative problem solving processes} into the RE process. To do this, a detailed study on selected \emph{creativity}  theories, techniques, training and tools has been conducted to see what can be adopted to improve RE and identified two creative problem solving processes that are poorly supported in RE. In [IEE16], a group of prevention strategies based on \emph{human errors} has been identified. In this work, 10 out of 21 prevention mechanisms reported by practitioners were related to preventing \emph{human errors} through changes to RE practices. 

[IEE20] describes an approach of using an iterative review process as the solution for the \emph{cultural} effects. It proposes templates and professional technical writing training that it claims will improve RE processes so that the practitioners can understand requirements in better way from a \emph{cultural} perspective. A set of strategies have been presented in [IEE25] with the idea of identifying highly \emph{creative} potential based strategies. The authors show that these are more effective than the use of lower creative potential-based strategies in prompting a worker to produce novel ideas in RE. Apart from proposing a theoretical framework, study [ACM05] has designed strategies related to \emph{motivation}. The authors claim their strategies may be adaptable, flexible, pragmatic and effective to address motivational issues.

Regarding \emph{communication} issues, [SP01] has provided guidelines to manage communication challenges during RE process as the key solution of the study. Meanwhile [SP16] has identified key causes that affect proper team \emph{communication} and they provide practical strategies to reduce the issues. These inclde approaches such as team members should be ideally \emph{distributed} between countries that have smaller \emph{time differences} to minimise \emph{communication} difficulty. In addition, \emph{distributed teams} should be provided with video, voice and/or text \emph{communication} options to address their communication challenges. In the study [SB05], a detailed guidance has been presented which focuses on identifying \emph{human values}. A taxonomy and questionnaire based approach has been used a support teams in identifying key values.

 \renewcommand\labelitemii{$\square$}
\begin{itemize}
    \item[$\square$] \textbf{Revealing effects of human aspects on RE}
\end{itemize}

Some studies have focused on revealing the challenges or effects of human aspects on RE as their main solution. For example, [IEE02] has revealed that \emph{inadequate communication} is the major issue. This creates major challenges when managing requirements across multi-site organizations. Similarly,  \emph{geographic distribution} has a significant impact on the \emph{collaboration} between the groups involved in the negotiation of requirements in a diverse environment. Study [IEE15] identified the effects that \emph{cultural differences} and \emph{personality characteristics} have on dynamic role allocation. They provide a detailed description of how \emph{group dynamics} are related to role allocation. Moreover, [IEE04] has recognised which \emph{personality traits} are more suitable for the people who contribute to each phase in SE. This includes identifying what they claim are the expected qualities for each phase including RE process.

According to [IEE14], requirements engineer \emph{domain knowledge} has a small but statistically significant effect on the effectiveness of the RE elicitation process. The same study has also identified that the \emph{expertise} of the interviewee is a more significant factor during requirements elicitation than the analyst's domain knowledge, and this has much more influence in final results in RE. Table \ref{tab: revelaed challenges/effect} present further details of the studies that have revealed the challenges/ effects of human aspects on RE as their solutions.\\

\begin{table}[]
\centering
\caption{Solution: Revealed challenges/ effect on RE}
\label{tab: revelaed challenges/effect}
\resizebox{\columnwidth}{!}{%
\begin{tabular}{@{}cll@{}}
\toprule
\textbf{Paper ID}         							 	& \multicolumn{1}{c}{\textbf{Human Aspects}}                                         		& \multicolumn{1}{c}{\textbf{Revealed challenges/ effect on RE}}                                \\ \midrule
\multirow{3}{*}{\begin{tabular}[c]{@{}l@{}}IEE02 \\ IEE11\end{tabular}}        		& \begin{tabular}[c]{@{}l@{}} \\ \\ Communication \\ and\\ Geographic \\ Distribution \end{tabular}         & \begin{tabular}[c]{@{}l@{}}Inadequate communication creates challenges when managing\\  requirements across multi-site organizations\end{tabular} \\
                              								                 &                         					       	& \begin{tabular}[c]{@{}l@{}}Impact on the collaboration between the groups involved \\ in the negotiation of requirements in a  diverse \\ environment\end{tabular} \\ \\
                             							 	&												&  \begin{tabular}[c]{@{}l@{}}Strong influence  on both individual and team behaviour, \\ especially when working on distributed, global software development \\ projects and multinational environments.\end{tabular} \\ \midrule
                               
\multirow{2}{*}{\begin{tabular}[c]{@{}l@{}}IEE17 \\ SP07 \\ WI04\end{tabular}}     	& Communication                                                                    		& \begin{tabular}[c]{@{}l@{}} \\Direct  effect  on  the  quality  of  the  requirements  elicitation  phase. \\ \\ Identified  seven  success  factors which cover approaches to better \\ managing aspects of collaboration, requirements understanding \\ and  communication in distributed team environments\end{tabular}       \\ \\
                              								& 				                                                     		& \begin{tabular}[c]{@{}l@{}}For an effective analysis of requirements, close interaction and \\ communication with customers are crucial.   \end{tabular}                                                            \\ \midrule
\multirow{1}{*}{IEE24}      								& \begin{tabular}[c]{@{}l@{}} Culture  \end{tabular}                                                   					& \begin{tabular}[c]{@{}l@{}}\\ Based on three dimensions in the Hofstede cultural model, 15 cultural \\ aspects  have been identified as having a significant impact on the RE process \\ considering Saudi Arabia and Australia. \\ \\ Also, identified that these aspects  may also be  applied to other cultures. \end{tabular}                                         
                                                                               \\ \midrule

\multirow{2}{*}{IEE15}    								& \begin{tabular}[c]{@{}l@{}} \\ Culture and \\ Personality \end{tabular}                          & \begin{tabular}[c]{@{}l@{}}\\Both the aspects effect on dynamic role allocation on RE/SE; provide detailed\\  description of how  group dynamics are related to role allocation      \end{tabular}                                            
                                                                                    \\ \midrule

\multirow{3}{*}{\begin{tabular}[c]{@{}l@{}}IEE04 \\ ACM20 \\ SP06\end{tabular}} 	& Personality										       & \begin{tabular}[c]{@{}l@{}}\\ Revealed which personality traits are more suitable for the people who contribute \\ to each phase in SE claiming  the expected qualities  for each phase \\ including RE  \end{tabular}                                         \\ \\
                               								&         											& \begin{tabular}[c]{@{}l@{}}Identify the influence of personality traits  of the practitioners in prioritization of \\ the needs and system requirements     \end{tabular}                                                              \\ \\
                               								& 											        & \begin{tabular}[c]{@{}l@{}}Based on their personal preferences, experience and personality, analysts \\ feel  comfortable or uncomfortable using elicitation techniques.                                 \end{tabular}    \\ \midrule

\multirow{1}{*}{IEE14}  								& Domain Knowledge									       & \begin{tabular}[c]{@{}l@{}} \\ Revealed that there is a small but statistically significant effect \\ on effectiveness of the RE elicitation related to analysts' domain knowledge  \\ that has much more influence in final results in RE  \end{tabular}                                                            
                               			                                                         					      \\ \midrule
                               
\multirow{2}{*}{IEE27}  								& \begin{tabular}[c]{@{}l@{}} \\ \\ Domain Knowledge \\ and \\ Communication \end{tabular}                 & \begin{tabular}[c]{@{}l@{}} \\ Revealed that domain knowledge is a contributing factor in the delivery \\ of quality requirements      \end{tabular}                                                           \\
                               								& 												   & \begin{tabular}[c]{@{}l@{}} Revealed that limited interaction and insufficient communication result \\ in poor requirements       \end{tabular}                                                          \\ \midrule
                 									                                                               
\multirow{1}{*}{IEE18}  								& Gender										                & \begin{tabular}[c]{@{}l@{}} \\ Identified that effect of female participation in requirements related tasks as they are \\ often assigned more into coordinating, planning and tracking the execution \\ tasks  which reveals the gender bias in RE     \end{tabular}                                                                                         \\ \bottomrule

\end{tabular}%
}
\end{table}

\begin{table}[]
\centering
\caption{Evaluation methods used in the studies}
\label{TABLE 11: Evaluation methods}
\resizebox{\columnwidth}{!}{%
\begin{tabular}{@{}ll@{}}
\toprule
\multicolumn{1}{c}{\textbf{Evaluation Method}}          & \multicolumn{1}{c}{\textbf{Paper ID}}                                              \\ \midrule
Case studies                                            & \begin{tabular}[c]{@{}l@{}}IEE02, IEE12, IEE15, \\ ACM04, ACM19, SB05\end{tabular} \\
Comparison with other models/ tools/findings/studies    & \begin{tabular}[c]{@{}l@{}}IEE01, IEE05, IEE06, \\ WI03, SP07, SB06\end{tabular}   \\
Conducting empirical studies/ controlled experiments    & \begin{tabular}[c]{@{}l@{}}IEE05, ACM19, SP05, \\ SP17, SB07\end{tabular}          \\
Use prototype/ story boards/ pilot study                & \begin{tabular}[c]{@{}l@{}}IEE01, IEE15, IEE16, \\ SB05\end{tabular}               \\
Follow-up interviews                                    & \begin{tabular}[c]{@{}l@{}}IEE01, IEE24, \\ ACM08\end{tabular}                     \\
Questionnaires-users/ research team/ participants       & IEE03, SP07, SB05                                                                  \\
Observing research design \& findings                   & ACM01, SP08                                                                        \\
Direct feedback                                         & IEE19, SP03                                                                        \\
Used other techniques (eg: MDS)                         & SP08, WI01                                                                         \\
Developing hypothesis - evaluate results and answer RQs & IEE25                                                                              \\ \bottomrule
\end{tabular}%
}
\end{table}

\begin{itemize}
    \item \textit{Evaluation of the solutions}
\end{itemize}

25 primary studies have used a variety of evaluation methods to evaluate the proposed solutions of their studies. These include the evaluation of a proposed prototype, theoretical framework, or developed model. As shown in Table \ref{TABLE 11: Evaluation methods}, the two highest number of studies conducted an evaluation using case studies (6 studies), and comparison with other methods/tools/findings/studies (6 studies). The results of the studies (eg: a new proposed model) were evaluated using more case studies or real world scenarios to identify the practical issues of using the model, and a comparison was carried out with other methods/tools/studies currently used in industry. Another common evaluation method was conducting empirical studies or controlled experiments, done in 5 studies. Using prototypes, story boards or pilot studies (4) and follow-up interviews  (3) are two other types of evaluation methods that have been used in several studies. Questionnaires with participants (3), observing research design and findings (2), and receiving direct feedback from users (2) are used by a few studies.
We have identified that in the rest of the studies (49 studies), various  types of solutions have been proposed. Most of these solutions are based on theoretical frameworks or models and they still need to be evaluated in future studies. From that, 40 studies have mentioned evaluation as the next step of the research whereas 9 studies have not mentioned it in the paper. \\

\begin{itemize}
    \item \textit{Limitations of the studies}
\end{itemize}
 
We have identified that many of the selected primary studies have similar kinds of limitations. We have categorized the limitations of the studies into 4 main groups -- evaluation results limitations, limitations with chosen methodology adopted, limitations with regard to study participants, and limitations with regard to focus area of the study.  
 \renewcommand\labelitemii{$\square$}
\begin{itemize}
    \item[$\square$] \textbf{Limitations in evaluation of the results}
\end{itemize}
This is one of the most common limitations in studies. We have identified that the limitations in evaluation mainly depends on the final outcome of the studies. In studies [IEE01], [IEE02], [IEE12], [IEE27], [ACM09], [ACM19] \& [SB05], it has been explicitly mentioned by the authors that their evaluation is limited, often to just one case study. They state that more validation of the outcomes with a larger number of case studies is needed. It is also often suggested that the case studies used are limited and should be from multi-site organizations, different software domains and different software teams, so that the proposed solutions can be shown to generalize. In [IEE03], [IEE06], [IEE13], [IEE15] \& [SP02], the evaluation of the solution models has been limited to the research team or direct participants in the study. Hence, these models are still to be evaluated with real software development scenarios incorporating industry practitioners to identify practical capabilities and limitations of these proposed solutions. A few studies, such as [IEE23], [SP04], [SP06] \& [SB01], only propose theoretical solutions. They completely lack any evaluation of their proposed solutions at all.

 \renewcommand\labelitemii{$\square$}
\begin{itemize}
    \item[$\square$] \textbf{Limitations in methodology used}
\end{itemize}
Limitations in methodology have been identified particularly for the data analysis methods and data extraction methods used in the studies. In [IEE01], it was mentioned that there is a consistency issue in their data analysis process. In [IEE05], there is a limitation in experience and cognitive capacity in analysing qualitative data, so that the methodology is only applicable when user logged data is readily available. Meanwhile, in [ACM01] \& [ACM09], it was mentioned that their data analysis process is limited due to not using  statistical methods on the resultant data. Moreover, the study [ACM03] has a limitation regarding the generalisation of their results due to an unbalanced data set extracted in the study. [ACM11][ACM12] \& [ACM13] studies have faced limitations based on the methods used for the data collection. These include using only a semi-structured interview, collection of RE related job ads, and use of a Q-sort method, all that effect the quality and completeness of the collected data sets. 

 \renewcommand\labelitemii{$\square$}
\begin{itemize}
    \item[$\square$] \textbf{Limitations in participants} 
\end{itemize}

Because of the nature of research on human aspects impacting the RE process, a very common study limitation is in regards to the participants used in the studies. This directly effects both the results and conclusions made from the studies and any possible generalisation of the study proposed framework, method or tool. This limitation can vary greatly based on the number of participants and the types of participants. In studies like [IEE03] and [IEE09], only research staff or academic professionals from institutes have been used as participants. This limits both the number of participants and whether they actually represent RE practitioner characteristics. In many studies, including [IEE11], [IEE18], [ACM04], [ACM10], [ACM18], [ACM20], [SP03], [SP07], [SP11], [SB04] \& [SB06], the participants were limited to student groups such as undergraduate students, postgraduate students, SE specialized students or non-IT students. This will also make an large impact on the generality and applicability of the final study results. There is usually a major difference between student experience, expertise, time commitment, and other aspects and actual IT professionals, particularly experienced requirements engineers.

Studies [IEE04], [IEE20], [WI03], [ACM03], [ACM09], [ACM13] \& [SP10] have included software industry practitioners as participants, but the number of software professionals who participated for their studies was very small (8-36 range). This means that the representativeness of these participants is limited and  generalisation of any study results becomes a major problem. Meanwhile some studies have faced difficulties due to low number of participants because of the type of the research study. The in-depth knowledge of the improvisation theatre technique of participants [IEE19], geographic location of the participants [ACM05][SB03], experience of the participants [ACM07] and physical issues of the participants [SP08], are some of the other reported issues that studies faced. All of these have  resulted in both a limited number of study participants and a limited diversity of study participants.

 \renewcommand\labelitemii{$\square$}
\begin{itemize}
    \item[$\square$] \textbf{Limitations in focused area of the study}
\end{itemize}
Some limitations were found based on the research area of the particular studies that are related to RE. These limitations can be based on the human aspects that were focused on, the considered RE phases in the study, and the selected organizations or countries for the study. In [IEE21], the study has only considered personality as a aspect that affects creativity. As a result the identification of the actual effect of personality on the RE process was limited. The study [ACM15] has only focused on the effects of personality, [SB03] was limited to personality and attitude, and [SP06] was about only the effect of communication issues. [SP15] was limited in its focus area to three human aspects; motivation, communication and domain knowledge. 

Studies including [IEE06][IEE16][SP05][SP12][SB02] \& [SB07] have limitations based on the focused RE phases. In [IEE06] and [SB07], the study focused on RE issues related to the global software development domain. [SP05][SB02] and [SB07] have only considered human aspects impacting the requirements elicitation phase. The study in [SP12] was about the requirements analysis phase, not the whole RE process. In contrast, [IEE16] considered 4 different phases of RE.

Several primary studies face limitations where they are limited to particular organizations, countries or separate geographic areas. In [IEE24], the study was about identifying the effect of culture in RE and it is limited to only Australian and Saudi Arabia. Cultures in other countries may have significantly different impact on RE phases than in these two examples. [ACM06] has focused on the effect of motivation on RE and was limited to only Brazilian companies. Another study [ACM08] investigated the effects of emotions and motivations, but only conducted their research study in one particular organization. The authors admit that their solutions may or may not be applicable outside that one organization. The study [ACM10] was conducted regarding gender effects on RE and was based in the US. The authors state that their study results might greatly vary if it considered worldwide women participation in RE. There are studies like [ACM12] and [ACM16] which are conducted based in one particular city. The study [ACM12] is limited to Brasilia and [ACM16] is limited to New York city.\\  

\begin{itemize}
    \item \textit{Suggested future work areas of each study}
\end{itemize}

All of our 74 primary studies have mentioned various types of future work. Based on the different directions for this future work, we categorized these into 4 main areas that we think will be helpful to inform future studies.

\renewcommand\labelitemii{$\square$}
\begin{itemize}
    \item[$\square$] \textbf{Validate or improve the proposed solution}
\end{itemize}

Many of the studies that proposed a new or modified model/framework as their solution also mentioned that these models/frameworks should be validated more thoroughly. For example in [IEE01], it was suggested to explore and investigate more on developed psychological-driven models. [IEE02] considered validation on the developed model with other multi-site organizations. In [IEE10], [IEE12], [IEE13], [IEE19], [ACM07] and [SP06] one of the key future work items suggested is to test and validate the presented solutions using real case studies. This will benefit in evolving the concepts and guide future researches.

\begin{itemize}
    \item[$\square$] \textbf{Extending the research based on current findings}
\end{itemize}

Another key suggested area of future work is extending the research presented, based on the findings of the current studies. Here, the studies have suggested various ways of extending their research based on limitations or gaps identified from the study. Replicating the same study in a different domain is suggested in many, including [IEE24] [ACM01] [ACM06] where the domain studied limits the findings. Increasing the number of participants of the studies and the collected data set is a common suggestion, for example suggested in [IEE11] [WI02] [SB01] [SB06]. 

\begin{itemize}
    \item[$\square$] \textbf{Investigate a new or related research area}
\end{itemize}

Investigating a new or related research area is a prominent future work suggestion mentioned in the primary studies. For example, study [IEE01] has examined the effects on emotions, motivations and human values. The authors suggest next to focus on other various types of psychological aspects. [IEE12], [IEE13] \& [IEE14] studies also suggest next to focus on other human aspects, apart from those aspects that each study has already considered. Moreover, the study [IEE17] suggested to identify a more systematic way of addressing RE issues. [ACM07] suggested to consider the whole RE process rather than one phase of it in future studies. 

\begin{itemize}
    \item[$\square$] \textbf{Develop a new model/framework based on the findings}
\end{itemize}

As some studies have only focused on proposing theoretical models or guidelines as the solutions, the developing of the proposed models were taken into considerations as future work. For example; [IEE02] suggested developing of an integrated RE tool that addresses all the identified communication and knowledge management challenges in the current study whereas [IEE24] suggested to develop a framework that describes the influence of culture on RE process identified in the current study. However these future work areas are differ based on what human aspect(s) have been discussed in the studies and it helps to identify key gaps in the area of identifying the effect of human aspects on RE.\\

\cornersize{.2} 
\ovalbox{\begin{minipage}{7.5cm}
\small
\textbf{Answers to RQ2:} A variety of human aspects have been studied to date. The majority of primary studies have focused on investigating just one human aspect by itself. Considering individually studied human aspects, the most studied aspect is communication issues (7 studies). The majority of the studies used existing domain models which tend to originate in the psychology research domain. 45\% of studies include a proposed solution to understand or address human aspects during RE. These range from theoretical models to practically applicable models, guidelines and tools. Key limitations of many studies to date include their evaluation, methodology and focus area.
\end{minipage}}

\subsection{What RE phases are most impacted by human aspects, and what are the relationships between different human aspects that affect these RE phases? (RQ3)?} \label{4.3.3 (RQ3)}
\begin{itemize}
    \item \textit{ Most affected Requirements Engineering phase(s) by human aspects}
\end{itemize}

Only 18 out of 74 primary studies focused on identifying the most affected RE phase caused by human aspects. The other studies considered the overall RE process as a whole, or RE as one phase in SE as a whole. These 18 papers focused on requirements elicitation, requirements analysis, requirements specification, requirements validation phases. We were unable to find any studies that have discussed the effect of human aspects on the requirements management phase. 

\begin{figure}
  \includegraphics[width=\linewidth]{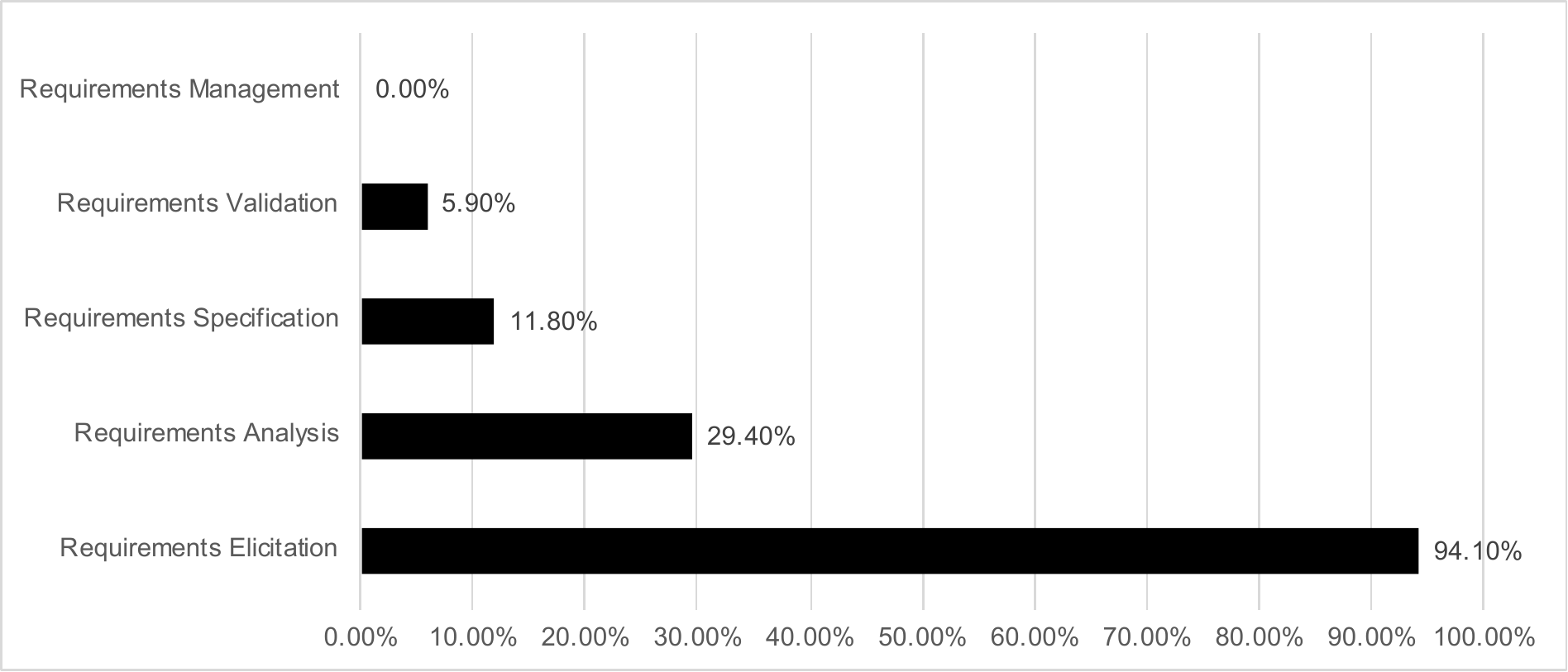}
  \caption{Research studies showing how human aspects impact on different RE phases}
  \label{Figure11:REphases}
\end{figure}

As shown in Figure \ref{Figure11:REphases}, requirements elicitation has been shown to be the most affected RE phase by human aspects. 8 studies have specifically discussed the effects on the requirements elicitation phase due to various human aspects. The majority of the studies were about the effects due to communication issues, where the effect has been identified individually [SP01],[SP05] \& [SB02] as well as combining other aspects. These include culture \& communication issues [SP02] \& [SB07], personality \& communication issues [SP06] or  geographic distribution \& communication issues [SB07]. The other two are about identifying the effect of human values [IEE06] and domain knowledge [IEE14] during the requirements elicitation phase.

Effects of human aspects on the requirements elicitation phase have been studied along with several other requirements engineering phases. Requirements elicitation and requirements analysis were studied in [IEE12][SP12]. The human aspects that were focused on were emotions, motivation, physical issues and domain knowledge. Requirements analysis and requirements specification phases are considered together in [IEE11]. This studied the effects of culture \& gender. These two phase have also been considered with requirements elicitation in [IEE01] and [IEE02], that explored the effects of motivation, emotions, human values and communication issues  on these three phases of RE. Only one study has focused on the requirements validation phase [SP03], and they specifically looked into how communication issues impact on the requirements validation phase. Though it has been identified that the most affected RE phase is the requirements elicitation phase, it has not been clearly identified that which human aspects impact it the most. \\

\begin{itemize}
    \item \textit{ Relationship between various human aspects that impact RE}
\end{itemize}

From our selected 74 primary studies, only 31\% have stated that there is a relationship between various human aspects and then tried to identify it. The relationships between motivation \& emotions, culture \& communication, personality \& motivation, personality \& human values and personality \& emotions are the most discussed relationships among the studies. [IEE01], [IEEE10] and [IEE12], explored whether emotions and motivation share a similarity, as both serve to define relations between individual and environment, and are tightly linked to actions taken. Moreover, it was identified that emotional reactions to scenarios may indicate motivational problems. This can lead to detrimental effects on the analysis of  requirements. 

Culture \&  communication issues seem to share another important relationship. According to studies [IEE02][IEE17]][SB07], cultural diversity leads to communication issues by creating a lack of awareness about the work.  This is most challenging in requirements elicitation, specification and validation phases. The relationship between personality and motivation has been discussed in [IEE12][IEE13][IEE26] and [SB05] studies, where it is mentioned that both psychological constructs relate closely to each other in an individual context. People's actions are assumed to be influenced by their personality which will also affect their motivation. Hence, people's personality dimensions or traits are related to motivation and higher motivation leads to better requirements analysis. 

Personality has another two relationships with human values and emotions. Considering personality and human values, the studies [IEE21] and [SB04]  focused on creativity as a human value and tried to identify the relationship between personality and creativity. They claim that personality dimensions also impact on an individual's creativity where diversity of personalities in a team improve the amount of creativity of the team. Unlike some of the previous human aspect combinations above, the relationship between personality and emotions is said to be an indirect one [SB01][SB06]. Individual personality influences work preferences and people's emotions also influence their work preferences. Hence, personality and emotions have an indirect relationship that will effect on the quality of the software product. 

Apart from these human aspect relationships, the studies have identified a variety of connections between human aspects such as human values \& motivation [IEE01], culture \& geographic distribution/time zone [IEE02], gender \& culture [IEE11], emotions \& gender [ACM18] and communication \& geographic distribution/time zone [SP16]. However all state that more detailed studies should be conducted to better identify these relationships.

\begin{itemize}
    \item \textit{Categorization of the most important (combination of) human aspects related to RE}
\end{itemize}
All the extracted primary studies have considered at least one human aspects and its impact on RE. Though the studies mentioned the importance of the human aspects and how it  impacts on RE, no research has identified the most important human aspects or the combination of aspects critical for RE. As shown previously in Figure \ref{Human_Factors}, we categorized human aspects into three main categories; \emph{individual human aspects, team related human aspects and technical related human aspects}. Many of the primary studies have focused on various combinations of these. The majority of the studies have considered communication issues (25 studies), personality (24 studies), emotions (17 studies), motivation (15 studies), human values (13 studies) and culture (13 studies). This implies that these aspects play an important role in RE.
However, the importance of different aspects may greatly vary and be unique. For example some  studies focused on two individual human aspects, motivation and emotions [IEE10] [IEE12] [ACM08]. Other studies have focused on individual and team related human aspects, culture and communication issues [IEE20]. Likewise, there are many unique combinations of human aspects and as this area is still an emerging area of research, more studies and experiments are required to determine which human aspects have greater importance in RE.

\cornersize{.2} 
\ovalbox{\begin{minipage}{7.5cm}
\small
\textbf{Answers to RQ3:} Few studies (18 out of 74) to date have focused on identifying the most affected RE phase caused by human aspects. The rest of the studies considered human factors and the overall RE process, and many consider other SE phases as well. The requirements elicitation phase has been identified as the most affected RE phase by human factors, but it has not yet been clearly identified which human aspects impact it the most. Considering relationships between human aspects, studies have focused on few combinations and more detailed studies are required to better identify these human aspect inter-relationships. 
\end{minipage}}

\subsection{How do the identified human aspects affect the RE Process (RQ4)?} \label{4.3.4}

\begin{table*}[]
\caption{Human Aspects impact on RE and SE}
\label{tab:Summary of affects}
\resizebox{\textwidth}{!}{%
\begin{tabular}{@{}lllll@{}}
\toprule
\textbf{Category}       & \textbf{Human Aspects} & \textbf{Impact on RE}   & \textbf{Impact on SE}  & \textbf{Paper ID} \\ \midrule
\multirow{6}{*}{\begin{tabular}[c]{@{}l@{}} Individual \\ Human \\ Aspect \end{tabular}}
& Personality 
& \begin{tabular}[c]{@{}l@{}} Improve interaction, \\ Think out of the box and more socializing \\ Openness, and conscientiousness influence the novelty of  ideas \\ and agreeableness conscientiousness, and extroversion \\ influence the usefulness of ideas \\ Being Extrovert helps to communicate \\ with users and management \\Important for understanding stakeholder groups \\ and for individual-level requirements \\ when systems can be customised or configured\end{tabular} 
& \begin{tabular}[c]{@{}l@{}} Improve the decision making \\  Assigning preferred task based on the personality will lead \\ to perform the task better - right people to right roles in SE \\ Personality type of the people involve in the team, \\ group diversity averted the disruption \\ improve SE process\end{tabular} 	
& \begin{tabular}[c]{@{}l@{}}IEE04, IEE25\\ ACM02 \\ ACM15 \\ SB04 \\ SB05\end{tabular}\\ \\                                                    

& Emotions			
& \begin{tabular}[c]{@{}l@{}} Capture more requirements \\ Increase the corrective actions in elicitation and analysis \\ Leads to identify specific situations \\ and requirements successfully \\ Negative emotions affect user rejection, \\ reliability and stability of the requirements	\end{tabular}
& \begin{tabular}[c]{@{}l@{}} Affected by visceral processing based on the appearance \\ or “look and feel” of a product \\ Potential to have conflicts  - have to consider \\ more on team composition when forming project team \end{tabular}                								   
& \begin{tabular}[c]{@{}l@{}}IEE01, IEE13\\ IEE10 \\ ACM19 \end{tabular}\\ \\     

 & Motivation 		        
 & \begin{tabular}[c]{@{}l@{}} Capture more requirements , \\ Increase the corrective actions in elicitation and analysis, \\ Leads to identify specific situations and \\ requirements successfully, \\ Engage more in Requirements elicitation, \\  Important for understanding stakeholder groups \\ and for individual-level requirements \\ when systems can be customised or configured \end{tabular}       										
 &  \begin{tabular}[c]{@{}l@{}} Potential to have conflicts - have to consider more \\ on team composition when forming project team \end{tabular}   			
 & \begin{tabular}[c]{@{}l@{}} IEE01, IEE13, \\ IEE10, \\ IEE26, SB05   \end{tabular} \\  \\                                 					       
 
 & Human Values		        
 & \begin{tabular}[c]{@{}l@{}} Capture more requirements, \\ Make RE process more innovative, \\ Increase the corrective actions in elicitation and analysis \end{tabular}	
 & \begin{tabular}[c]{@{}l@{}} Effect the innovation and originality of the project, \\ Potential to have conflicts -have to consider more \\ on team composition when forming project teams \end{tabular}                             
 & \begin{tabular}[c]{@{}l@{}} IEE01, IEE02, \\ IEE13, \\ IEE23 \end{tabular} \\ 	\\	                                                                              
 & Culture			
 &  \begin{tabular}[c]{@{}l@{}} Better performance in SRA by multinational teams, \\ Helps to adapt to the situations (comfortable with \\ cross cultural situations) and perform RE process confidently, \\ Better understand requirements \end{tabular}                       & \begin{tabular}[c]{@{}l@{}} Can cause fear, mistrust or other social problems \\ which increase the failure risk \\ Differentiate the phases of software \\ development life cycle \end{tabular}							                    
 & \begin{tabular}[c]{@{}l@{}} IEE11, IEE24, \\ IEE20, \\ IEE08, SB08  \end{tabular} \\   \\                                                                                                                 
 & Physical Issues  		
 & \begin{tabular}[c]{@{}l@{}} Resistance to ask more questions - leads to lack of requirements \\ elicitation and apparently it will doom the project  \end{tabular}			& 								                                                
 & IEE09 \\ \\  \midrule                                                                                                        
                                                                         
\multirow{3}{*}{\begin{tabular}[c]{@{}l@{}}Technical\\ Human \\ Aspects \end{tabular}} 	 
& Domain Knowledge              
& \begin{tabular}[c]{@{}l@{}} Effectively conduct requirements elicitation activity, \\ Missing or incomplete requirements or wrong information collection \end{tabular}	       
&  Overall project failure  														 
&  IEE14, IEE19 \\   \\                                                        

& Task workload                 
&                             			   
& \begin{tabular}[c]{@{}l@{}} Have to revise the system to reduce or \\ eliminate overload, and  wait till human factors \\ experts to make suggestions \end{tabular}					& IEE03 \\ \\

 & Human Errors                  
 & \begin{tabular}[c]{@{}l@{}} Misunderstanding of requirements, \\ Quality of the requirements will be affected  \end{tabular}  									
 & Failure of software projects														
 &  IEE16, IEE22  \\ \\  \midrule                                              
 
\multirow{3}{*}{\begin{tabular}[c]{@{}l@{}}Team\\ Human \\ Aspects \end{tabular}}                                                     
& Communication issues           
& \begin{tabular}[c]{@{}l@{}} Help to derive real needs of stakeholders, \\ Help to have better interaction between users and team \\ Improve the quality of requirements, \\ Increase the performance between client and analysts \\ during requirements validation, \\ Impact strongly on productivity and integrity \\ in requirements negotiation \end{tabular}		& \begin{tabular}[c]{@{}l@{}} Reduce developers rework \\ Increase less amount of errors in later phases \\ of Software development \end{tabular}					
&  \begin{tabular}[c]{@{}l@{}} IEE19, IEE20, \\ IEE27, SP03,\\ SP07, SP12, \\ SP15 \end{tabular} \\  \\     

& Geographic Distribution/ Time Zone  
&  \begin{tabular}[c]{@{}l@{}} Unsuccessful face-to-face interview \\ and brainstorming sessions limit the creativity of RE \\ Impact on requirements gathering, negotiation and specification \\  Lack of a common understanding of requirements  \end{tabular}				& \begin{tabular}[c]{@{}l@{}} Cause lack of effective communication \\ and lack of team work \\ Ineffective communication may increase cost \\ and lead to system and project failures. \end{tabular}  												
& \begin{tabular}[c]{@{}l@{}}  IEE06, IEE02, \\ SP01, SP02 \end{tabular} \\ \\                                                      
& Knowledge Sharing                   
& \begin{tabular}[c]{@{}l@{}} Reduce the trust level and ability to share work artifacts \\ during RE  \end{tabular}			
&			
& IEE02 \\ \bottomrule 
\end{tabular}}%
\end{table*}

The majority of the selected 74 primary studies are focused on identifying the impact of human aspects on the RE process. The rest point out the impact on the overall SE process or the product. Several studies discuss how people's personality affects  the RE process, such as how being extrovert improves communication and interaction with the users, and how openness and conscientiousness influence the novelty of ideas which will improve the quality of the requirements gathered [IEE04][IEE25] [ACM02]. Some studies discuss this personality affect not only in RE, but also in SE. For example, [ACM15] discuss the importance of assigning right people to right roles in SE based on their personality which will lead to perform the task better. In [IEE02], [IEE06] studies, lack of common understanding of the requirements has been identified as the affect of Geographical distribution which will impact on requirements elicitation, negotiation and specification phases whereas in [SP01], [SP02] studies identified that Geographical distribution may result in poor communication which increase the cost of the project, lead system and project failures. Table \ref{tab:Summary of affects} provides a summary of the affects of human aspects in RE/ SE separately considering the list of human aspects that we have synthesized from the reviewed studies (Figure \ref{Human_Factors}). 

Moreover, these affects of each human aspects have been categorized according to its nature, considering whether it has positive, negative or mixed impact  and following subsections focus on a detailed analysis of the effects of human aspects on RE based on the its nature.

\begin{itemize}
    \item \textit{Nature of the effect of the human aspects in RE Process; Positive or Negative}
\end{itemize}

Among the selected primary studies, 56.8\% report that they focused on identifying the impact of human aspects in RE process. The impact was categorized as positive, negative, both positive and negative or no impact. 

\begin{figure}
  \includegraphics[width=\linewidth]{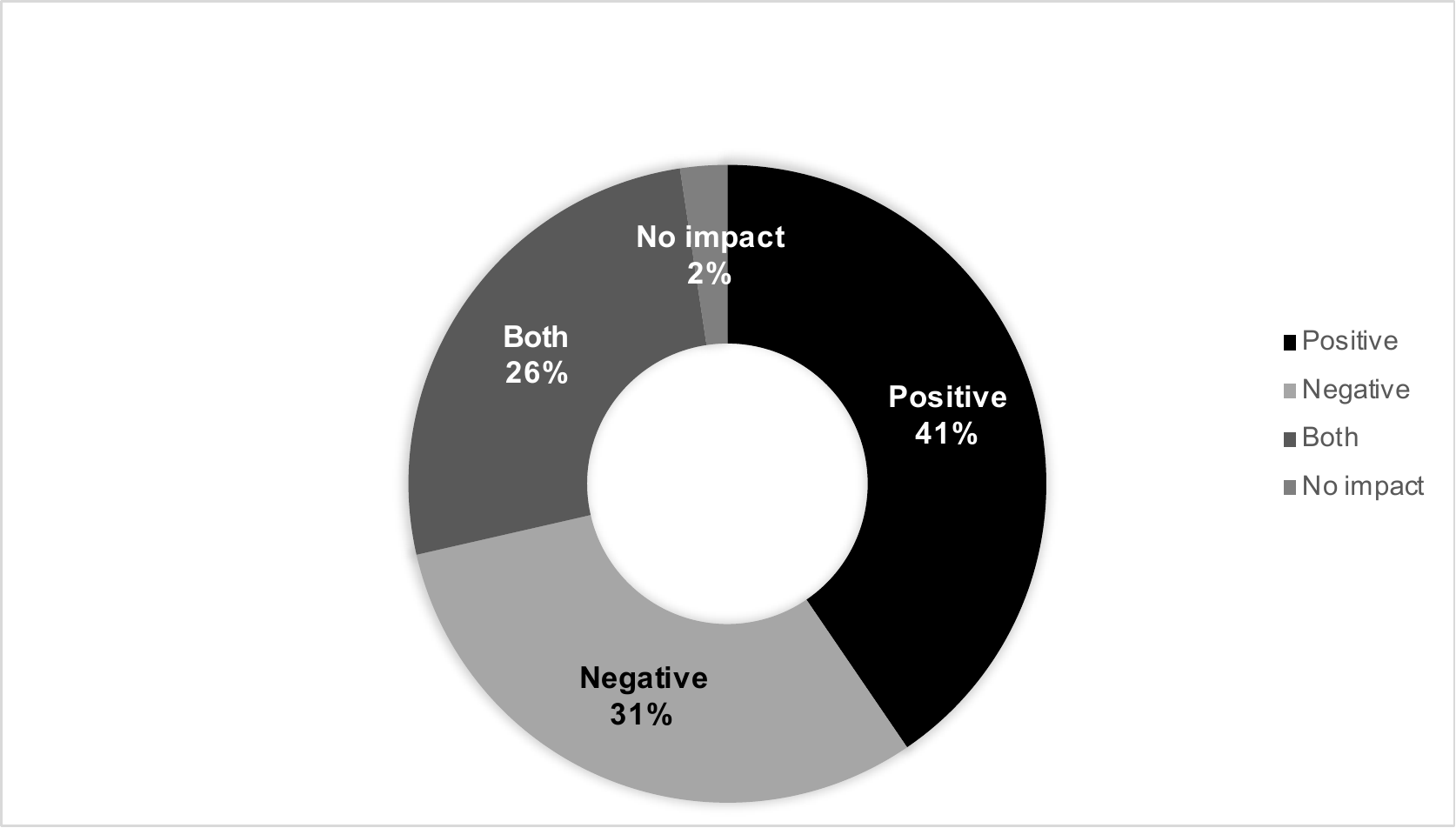}
  \caption{Nature of the impact of human aspect(s) on RE}
  \label{PositivNegative}
\end{figure}

Figure \ref{PositivNegative} shows that from these studies, 40.5\%  identified that there is a positive effect. 31\% of the studies determined that the nature of the effect of human aspects on the RE process was negative. Meanwhile, 26.2\% of the studies identified that there was both positive and negative effects from human aspects on the RE process. Only 2.3\% of the studies found that there was no impact of human aspects on the RE process. These suggest that a variety of human aspects positively or negatively effect the RE process.  Thus it is important to incorporate due consideration of positive supporting human aspects for the improvement of the RE process, while mitigating some negative impacts. 

\begin{itemize}
    \item \textit{Benefit of considering the Positive aspects in RE}
\end{itemize}

Some human aspects positively effect on RE process and may thus be beneficial for enhancing the RE process. Considering the studies that have reported positive effects of human aspects on RE, these benefits can be categorized based on following criterion.
 \renewcommand\labelitemii{$\square$}
\begin{itemize}
    \item[$\square$] \textbf{Improving the overall RE process}
\end{itemize}
Several studies report that human aspects such as personality, communication issues and culture can have a positive impact on improving the overall RE process [IEE19] [IEE20] [ACM02] [ACM11] [SP07] [SB04]. It was determined that in the RE process, people with extrovert personality have better interaction and communication skills with other stakeholders and clients. This will likely result in a better understanding of the requirements by better deriving the  needs of the clients. Hence, the RE outcomes will improve and eventually this will produce better software for the users and will reduce the amount of rework required in the project. 
\begin{itemize}
    \item[$\square$] \textbf{Improving quality in the RE process}
\end{itemize}

Communication issues and culture were found to positively influence  improving the quality of the requirements produced in the RE process [IEE11][IEE27][SP03]. Communication skills improve the quality of the interaction between analysts and clients. This results in clearer, more accurate requirements capture. Culture influences on diversity of thinking with multinational teams helps to elicit and capture better requirements.   
\begin{itemize}    
    \item[$\square$] \textbf{Making productive relationships among team i.e. the right people in the right roles}
\end{itemize}
It was found that human values, motivation, emotions and personality all impact on RE process by helping to form and support productive relationships among the software team [IEE13][ACM15]. These productive relationships can be built up among the team members including by identifying the right people to fill particular roles in the team. This can be done with the help of individual human aspects of team members. The authors claim that this better team formation and role structuring will result in increasing the performance of each SE phase including the RE process. 

\begin{itemize}       
    \item[$\square$] \textbf{Improving effectiveness of RE phases e.g. elicitation and analysis}
\end{itemize}
It was also reported that human aspects including personality, motivation, emotions, communication, human values and domain knowledge also have a positive impact on improving the effectiveness of RE phases [IEE01][IEE14][IEE16] [SP01][SP15][WI04][SB05]. It was shown that an individual's personality and motivation increase their engagement in the requirements elicitation phase. This then helps them to more effectively capture requirements. The use of effective communication also impacts on eliciting more accurate requirements. Meanwhile, requirements engineer domain knowledge improves the effectiveness of the requirements analysis phase as the requirements engineers have a better understanding of the requirements. Emotions and human values help requirements engineers to better empathise with and understand clients. This was shown to provide a beneficial effect on both capturing requirements and analysing them.   

\begin{itemize}   
    \item[$\square$] \textbf{Improving requirements engineers' decision-making, open-mindedness and confidence}
\end{itemize}
Another positive impact of human aspects that was determined in a few primary studies was that they can help to improve requirements engineers' decision-making, open-mindedness and confidence when engaging in the RE process. In [IEE04] and [IEE24], it was shown how personality and culture can both positively influence and improve requirements engineers' abilities. For example, When requirements engineers have an extrovert personality type, they are keen on achieving more effective communication with their clients, socializing with them, and thinking ``out of the box". These all help to improve their decision making and confidence when making RE-related decisions. Their cultural differences may also help them to adapt to different requirements engineering situations as being more  comfortable in them.  This can help them to be more open-minded when engaging in the RE process. 

\begin{itemize}   
    \item[$\square$] \textbf{Adding novel, creative performance and innovation to RE}
\end{itemize}
Studies [IEE02], [IEE23, [IEE25] and [SP04] show that some human aspects can help to improve novelty, creativity or inventiveness in the RE process. Openness and conscientiousness personality types can have a positive influence on the resulting novelty of ideas. Individual creativity can also help to make the RE process more innovative, and the authors claim this can help software organizations and their clients gain a competitive advantage. 

\begin{itemize}   
    \item[$\square$] \textbf{Reducing errors in RE and later phases in software development}
\end{itemize}

Reducing errors during RE is another positive impact due to human aspects such as improved communication, found in study [SP12]. It was identified that when there is less problematic communication issues between requirements engineers and clients and between requirements engineers and developers, there will likely be less errors in the RE process and later phases of software development using the requirements. Effort put into improving these communications is thus likely to provide a high payoff in terms of reduced errors and improved RE and software quality.\\


\begin{itemize}
    \item \textit{Effect of Negative aspects and the approaches to mitigate it}
\end{itemize}

Although the majority of primary studies that reported an effect of human aspects on RE mentioned various positive impacts, sometimes there are negative ones as well. This includes the studies that directly identify negative impacts of human aspects on RE, as well as the studies that have discussed both negative and positive impacts that they found. 
Some of the studies have suggested approaches that could be used to mitigate some of these negative effects. 

\textbf{Emotions} can be negatively effected  during RE and some mitigation strategies have been suggested to overcome their negative effects. In studies [IEE01], [IEE10], [IEE12], [SB04] and [SB05], it was reported that negative emotions directly affect user rejection, eliciting and analysing of requirements, the reliability and stability of the requirements, and they may increase the potential to have conflicts and impact on the outcome of the product.
These emotions may come from requirements engineers, clients, other developers or all three. Some mitigation approaches have been suggested for this impact in studies [IEE10] and [IEE12]. These include  considering more carefully team composition when forming project teams, and by analysing obstacles to developer motivations. Here,  emotional responses can be used to invoke appropriate motivations, while  potential negative emotions might be detected and converted to  more positive responses.

\textbf{Communication problems} is another major human aspect that can be effected negatively when it is not properly supported. According to several studies, including [IEE02], [IEE17], [SP01], [SP02] and [SB07], communication issues may result in several serious negative impacts in the RE process. It will directly impact on requirements elicitation, specification and negotiation with a lack of common understanding of requirements, missing or incomplete requirements, wrong information collection, and poor delivery of requirements. This often occurs where developers face difficulties in understanding what clients really want. This will result in client dissatisfaction, higher cost of the project, lower RE and software quality, and finally potential overall project failure. To mitigate these negative communication effects, the studies suggested introducing specific training that will reduce communication gaps between the parties, provide prior knowledge about the clients, using a communication check list guide  requirements engineers, and better considering and incorporating the communication human aspect when forming and managing RE teams [IEE02][SP02][SP03].  

\textbf{Culture} can also negatively effect RE, often together with problematic communication issues. Cultural differences can lead to miscommunication which will delay the RE process, reduce requirements quality, and add more work to other software team members. Moreover, it can cause fear, mistrust or other social problems which increases the risk of failure and slows progress of the project [IEE08][IEE24][SB07]. Together with culture, geographic distribution and time zone issues can also effect negatively on RE process. Geographic distribution and different time zones leads to difficulties in face to face discussions and brainstorming sessions, limiting  creativity in RE, as well as being more challenging for team communication. Moreover, it may lead to difficulties of dealing with large amounts of information from various sources, and obtaining, validating and refining requirements requires extra effort [IEE06][SP16][SB07]. According to these studies, the negative effects of time zone and geographic differences can be mitigated through engaging in more causal discussions prior to formal meetings, looking to change the distribution of team member roles between countries that have similar time differences, and using various tools e.g. the ShyWiki tool that has been proposed to help overcome distributed brainstorming problems.

\textbf{Lack of motivation, different human values, lack of domain knowledge, heavy task workload, human errors} and \textbf{restrictions on knowledge sharing} are other human aspects that can have negative effects on RE. Individuals with lack of motivation and different human values may have an increased risk of having conflicts. This may be mitigated by paying more attention to team composition when forming project teams [IEE12][IEE13][SP04]. Task workload is another aspect that may negatively impact on RE. With overloaded tasks, it is difficult to conduct proper RE processes. The tasks needs to be revised or reduced to overcome this problem. 

Due to \textbf{human errors}, requirements can be misunderstood and the quality of the requirements will be affected, resulting in failure of software projects. The studies suggested some prevention strategies to mitigate the effect of human errors, including designing a communication plan, creating dictionary/glossary of terms, and better knowledge transfer within the RE team so that human errors can be reduced [IEE16][IEE22]. \textbf{Lack of domain knowledge} and \textbf{restrictions on knowledge sharing} may lead to wrong requirements collection and missing actual requirements. This negatively impacts requirements elicitation and specification [IEE02][IEE19].  However, most of these studies have only identified the negative impact and suggested mitigation approaches. Further studies are needed to verify that these suggestions might actually be practical and work.

\cornersize{.2} 
\ovalbox{\begin{minipage}{7.5cm}
\small
\textbf{Answers to RQ4:} The majority of the studies focused on identifying some form of impact of human aspects on the RE process. These impacts have been identified as positive, negative, both positive and negative, or no impact. Many studies (40.5\%) identified that there is a positive impact. Some studies identified some negative impacts and many of these suggested some mitigation strategies as well. However, further studies are needed to evaluate these strategies in practical RE scenarios. 
\end{minipage}}

\subsection{Identified key research gaps} \label{4.7}

\begin{figure}
  \includegraphics[width=\linewidth]{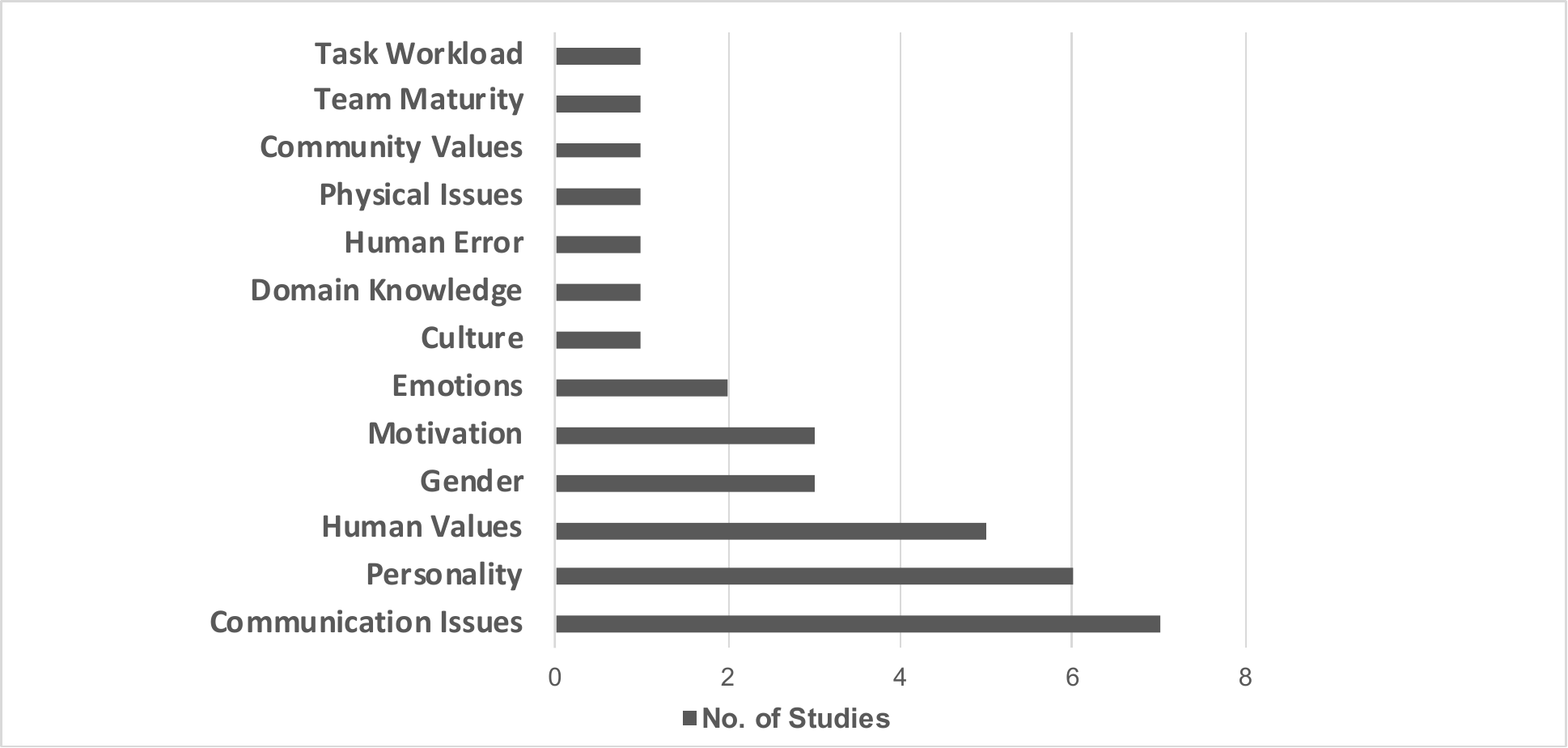}
  \caption{No. of studies on individual human aspects}
  \label{Individual aspects}
\end{figure}

As shown in Figure \ref{Individual aspects}, among the primary studies selected, 33  focused purely  on the effect of one human aspect, 19 studies were focused on combination of aspects, and the other 22 studies considered combinations of more than two human aspects in their studies. The individually most studied aspects were \textbf{communication issues (7 studies), personality (6), and human values including creativity (5)}. Gender, motivation and emotions have been investigated in 2 or 3 of the studies.  human aspects such as attitude, age, geographical distribution/time zone, self-management appear individually in one study each. 

Considering combinations of human aspects, most of the combinations were unique. The most commonly studied  combination was \textbf{emotions and motivation}, which was investigated in 3 studies [IEE10][IEE12][ACM08]. Combinations such as personality, human values and communication issues [SP10][SP12], personality and communication issues [SP06][SP07], emotions, motivation and human values [IEE01] [SB04] and culture, communication issues and geographic distribution [IEE17][SB07] were also considered in more than one study. All the other human aspect combinations have been studied once opening several possibilities of future research areas.  Personality \& emotions [SB01], personality \& attitude [SB03], culture \& communication issues [IEE20], culture \& geographic distribution [IEE08] and human values, communication \& culture [IEE02] are some unique combinations that have been considered only in one of the selected primary studies. 
    
When observing \textbf{personality}, almost all studies focused on the whole SE process, with RE just one phase considered [IEE25], [IEE05], [ACM02], [ACM04], [ACM15], [SB04]. As a result there are a lack of primary studies that directly focus on the effects of individual personality on the RE process, or sub-processes of RE. This contrasts with identification of  \textbf{communication issues} that have often specifically targeted RE phases, such as requirements elicitation [SP01][SP05][SB02], requirements validation [SP03], or the whole RE process [ACM11]. Hence, there is a gap in research to date in identifying how individual personality might effect RE phases, positively or negatively. 

\textbf{Gender} is another human aspect that lacks much research to date in terms of its impact on the RE process. Again most of the studies related to gender have focused on SE in general, including RE as one phase [IEE18][ACM10][ACM16]. Hence, there is a lack of studies that focus on potential gender bias situations or other gender issues and its effect in RE. \textbf{Motivations and emotions} are  two important human aspects where impact needs to be considered in the RE process. Several studies have identified that they have both positive and negative effects on the SE process. These two aspects have been investigated together in 3 studies, and 2 have specifically focused on their impact on RE [IEE10][IEE12]. Human aspects including \textbf{domain knowledge, human errors, physical issues, team maturity, attitude} have been considered related RE, but there are few studies regarding each aspect. Thus there is a clear possibility of valuable future studies focusing on each or combinations of these aspects.  

Another research gap we identified is that among the 74 primary studies, 75.7\% of them have not identified what is the most effected RE phase from the human aspects that they considered. This shows that though the studies focused on identifying the effect of human aspects in RE, the majority were unable to identify the most effected phase. Few studies have focused their research on one particular phase in RE. For example, in [SP01][SP05][IEE07][IEE26] and [SB02] the requirements elicitation phase was specifically studied. In [SP03], the requirements validation phase was specifically studied. This implies that there is a need for more studies that explain the effect of different human aspects on all RE phases. 

We have also identified that only 31\% of the selected primary studies have investigated relationships between different human aspects and then focused on identifying what these relationships are. Studies such as [IEE08] [IEE20] [IEE22] [IEE25] [IEE27] [SP09] [SP10] [SP12] and [WI02] have focused on more than one human aspects. However, they have not identified whether there is a relationship or not between these human aspects in RE. Hence, there is a need to better identify the relationships between various human aspects, so that a proper taxonomy of human aspect impact on RE can be designed.  Considering each human aspect in the studies, there are a variety of areas that researchers could usefully focus on in future studies.

\subsection{Threats to validity} \label{5.2}

A standard threat of all SLRs is selection bias. We have addressed this by conducting comprehensive searches using the most commonly used databases by other SLRs in the SE context. We also followed Kitchenham and Charters \cite{RN2753} guidelines closely to help in minimizing errors when conducting this SLR. However, due to the variety of search strategies that we had to use for searching different databases and limitations of lengthy search strings, we faced some challenges in finding all the possible related studies. To overcome this, we used several search strings for one particular database and checked the search string that gives most number of related studies. We conducted this step for each database to find out most number of related studies for our SLR. We also used forward snowballing from references to find any additional studies that may have been missed. There is however still a possibility of missing a few related studies that have been published after our paper selection process.

Another challenge we faced was how to minimize  inaccuracy when performing data extraction. For that, the data extraction process was conducted with the close supervision of both second and third authors of this paper. They were closely involved in deciding how and what data would be extracted, and conducting some duplicate data extraction and compared result to the first author, discussing differences and resolving via consensus to minimize bias in the data extraction process. The evaluation of the quality of selected papers was also challenging. As the majority of the papers have been published recently, we were not able to take into account the citation count as a reliable quality criteria. Hence, we have focused more on the content of the papers as well as the published venues of the papers to overcome this challenge. \par
Grouping of human aspects was another challenge that we faced during the data synthesis phase of the SLR. Since we are considering all human aspects reported in the selected primary studies, there were a variety of terms used for the same human aspects and categorization of these aspects was a challenging task without a standard taxonomy of human aspects in the SE context. To overcome this issue, we decided to do the groupings based on the definitions of each aspect which have been already used in SE context.  We have focused on RE or SE papers that consider RE as a one phase of it and, excluded the papers that focus on other areas of SE though they have focused on human aspects. Hence, there is a challenge of finding all the appropriate studies that may be missed due to not using all possible search terms. For example; we used a wide variety of human aspects and RE terms to get most appropriate studies, but there may be any related studies in e.g. HCI and Design domains that do not use the term 'requirements'. We did look fo such papers, including these terms in our searches, but we were unable to find any human aspect related studies impacting RE. This can be considered as another area of further study using a wider range of terms related to these non-SE domains. 
\par Not having sufficient details in the primary studies about target application domain,size, granularity, team size, and organization size is another challenge that we faced during data extraction and analysis. It is likely that these characteristics have a potential to be influenced by various human aspects, or might not be impacted at all. Many of our primary studies did not provide such details as they have generally focused on SE/RE teams and processes. Hence we were unable to extract and synthesize them to a comprehensive extent.


\section{Recommendations} \label{section 5} 

As requirements engineering (RE) is one of the most crucial processes in SE, improvements in the RE process will improve the whole SE process \cite{RN1591}. Hence, it is vital to consider human aspects related to RE which have not yet been paid much attention to in SE research to date. Based on the findings of this SLR, reported future work and identified key research gaps, we have identified several key challenges in the area of studying the impact of human aspects on RE. We frame these below as a set of recommendations to the SE research community for further research in human aspects on RE.

\textbf{1. More studies are needed directly focused on the effects of human aspects on RE:} Among the selected 74 primary studies, only 35 studies focus on specifically RE, whereas the rest considered the SE process in general, RE being discussed often as a minor part. More RE-focused research on investigating the effects of human aspects is needed. 

\textbf{2. More practical guidelines and recommendations:} Most primary studies to date investigating the impact of human aspects on RE provide theoretical or academic models, strategies or prototypes. Few have focused on providing working models that were  trialled in the software industry. Many mention validation of these theoretical solutions as  future work. Studies providing a working model, tool or set of guidelines that can be practically used in real-world RE processes by software practitioners would be very helpful.

\textbf{3. Real-world, representative evaluations:} Many studies used academics and students, and those running industry based trials were mainly restricted to a single team, project, company, or country. New studies should address these limitations by validating their proposed solutions with more representative and larger scale, more diverse situations.

\textbf{4. Some human aspects may need more study in terms of their impact on RE:} 33 primary studies  focused on only one human aspect. Among these, communication issues and personality differences were the two human aspects considered individually. However, communication issues has been studied mostly related to RE, whereas personality has been studied mostly related to SE in general. Only a few other human aspects were studied in more than a couple of studies, including emotions, motivation, human values, geographic and time zone differences. Many human aspects were investigated only in a very few studies, as shown in Figure \ref{Individual aspects}, and their impact on SE generally and not specifically RE or RE phases.

\textbf{5. More studies are needed that consider multiple human aspects impacting on RE:} Very few primary studies consider more than one human aspect. Only a few combinations of human aspects have been investigated in a single study. We observed that the emotions and motivation combination is the most studied combination -- in only three studies. There is an opportunity for more studies on the effects of different combinations of human aspects on RE. 

\textbf{6. More investigation is needed on what are actually the most influential human aspects on RE, both positively and negatively:} All of our primary studies investigated some factor relating to the impact of human aspects on RE. However we could find no study that directly discussed identifying what are the \textbf{most} influential human aspect(s) on RE. Some human aspects may impact RE but not much. Others might have a great influence, in general, in combination, or in certain situations. 

\textbf{7. Investigating which phases of RE are more impacted by human aspects:} Only a quarter of the primary studies focused on identifying the most affected RE phase by human aspects, the majority focusing on requirements elicitation. Impact of human aspects on other RE phases are much less investigated. 
Conducting studies which try and identify the impact of human aspects on different phases of RE, and which phase(s) are more highly affected by which human aspects, would be valuable.

\textbf{8. Identification of the relationships between human aspects and their effect on RE:} Only 23 (31\%) of the primary studies identified that there is a relationship between the different human aspects considered in the studies and conducting RE. These human aspects can positively or negatively impact the RE process. Determining what the effects of different human aspects and combinations of human aspects are on RE is still an emerging area of research. There is a need in future research projects for methods to both capture what these human aspects are, identification of the actual positive and negative effects of these human aspects on RE, and investigation of tangible ways we can improve the RE process by incorporating due consideration of these  human aspects. 

\textbf{9. Investigating the influence of the target application domain, team size, organization size, etc:} 
We have not directly analysed the influence of the domain studied in the primary studies, the team or organisation sizes, etc. Also, we identified that there is lack of sufficient data in many of the primary studies to be able to extract and synthesize all of these within this study. Hence, there is a potential for another study that focuses on the influence of various human aspects based on different software project domains and team and organisational characteristics and their impact on the RE/SE processes.

\textbf{10. Need of a more comprehensive and agreed taxonomy of human aspects:}  When we identified our set of human aspects that have been studied to date, we tried to categorize them into broadly different kinds of human aspects. While doing this, we identified that there is a need for a more comprehensive and agreed taxonomy of human aspects in SE in genera, including wide range of keywords representing these. Building this taxonomy is another broad research area and we have begun a project investigating this aspect which can be incorporated into future studies.

\section{Conclusion} \label{section 6}
To conduct this SLR, 74 primary studies related to human aspects of RE process were selected and the human aspects were categorized into three areas namely; \emph{individual, technical}, and \emph{team}. The majority of the primary studies have focused on individual related aspects. Among the studies, 35 focused purely on the RE process and the rest focused on SE in general, but including RE as a phase of it. Among the set of human aspects, \emph{communication} issues (7 studies), \emph{personality} (6), \emph{human values} (5), \emph{gender} (3), \emph{motivation} (3) and \emph{emotions} (2) and \emph{culture} (2) are the individually most considered aspects. 
\emph{communication} issues were the most widely studied aspect related to RE including considering the impact of communication on different phases of RE. Even though \emph{personality} was the second most studied human aspect, it was not studied specifically focusing on RE, but in terms of SE in general. A number other human aspects impact on RE have only been considered in single studies to date, and some not at all.
These may benefit from investigation in future studies, individually and/or in combination. 

The main aim of this research was to identify how human aspects affect the RE process by answering 4 research questions (RQs). The first RQ revealed the key motivation, goal and objectives of conducting each of the primary studies on the effect of human aspects in RE. The second RQ targeted at identifying the current status of research studies in this area, considering what human aspects have been studied to date relating to the RE process, what kind of methodologies have been used to study them, how their evaluations have been performed, and overall limitations and potential future work proposed by each of the studies. Our third research question tried to identify the human aspects that have been shown to be the most influential on the RE process, and the relationships (if any) between different human aspects and their impact on RE. We found that so far there have been no studies conducted to identify the most influential aspect(s) on the RE process and how these human aspects inter-relate to impact RE. Our fourth RQ focused on identifying the positive and negative impacts of human aspects on the RE process that have been found to date. The majority of the studies found a positive impact (40\% of studies) and 32\% found a negative impact. Some studies found both positive and negative impacts, or no impact, of different human aspects on the RE process.

The findings of our SLR will be beneficial for understanding the effect of different human aspects on RE process.  The research community can focus on specific under-researched human aspects, aspect combinations, RE phases, developing practical guidelines and tools, and conducting studies with industry to better identify their impacts and provide better solutions. By understanding these potential  effects of human aspects on RE, software industry practitioners can be take due consideration of them when forming and managing teams, conducting the RE process and its phases, and in leveraging positive effects to improve RE while mitigating any negative ones.


%


\ifCLASSOPTIONcompsoc
\section*{Acknowledgments}
Hidellaarachchi and Madampe are supported by Monash Faculty of IT PhD scholarships. Grundy is supported by ARC Laureate Fellowship FL190100035 and this work is also partially supported by ARC Discovery Project DP200100020.
\else

\fi

\ifCLASSOPTIONcaptionsoff
  \newpage
\fi

\appendices
\section{Data extraction form fields} \label{A}
\begin{footnotesize}
\begin{enumerate}
  \item Paper ID.
  \item Paper Title.
  \item Authors of the Paper.
  \item Published Year.
  \item Venue (Name of the Journal/ Conference published).
  \item Authors' Affiliation.
  \item Type of Study: Journal publication/ Conference paper/ Book chapter/ Other.             
  \item Source Type: ACM/ IEEE/ Springer/ Wiley.
  \item What is the aim/ motivation/ goal of the study?
  \item What are the major Keywords of the study?
  \item Abstract of the study?
  \item What are the Research Questions addressed in the study?
  \item Subjects used in the study: Professionals or Undergraduates (Requirement Engineers/ Stakeholders/ Clients/ Students)?
  \item What are the human aspects that are considered in the study. (Personality/ Culture/ Emotions/ Human values/ Motivation/ Domain knowledge/ other)?
  \item What phases of the RE are considered in the study (Elicitation/ Specification/ Analysis/ Validation/ Management)?
  \item Does the research identify the most affected RE phase by human aspects? (Yes/ No)
  \item If Yes, What is the most affected RE Phase(s)?
  \item Does the study use any existing domain models related to human aspects? (Yes/ No)
  \item If yes, what are the existing domain models used in studies to identify human aspects?
  \item Method used in the study(s)? (Case studies/ Interviews/ Modelling/framework/ Document analysis/ surveys/ other)
  \item Is the study conducted based on academia or industry? 
  \item The number of participants used in the study.
  \item What type of data analysis used in the study? (Qualitative/ Quantitative/ Mixed)
  \item What are the main limitations of the study?
  \item What are the key research gaps/ future work identified by each study?
  \item Does the research focus on identifying the relationship between different human aspect(s)? (Yes/ No)
  \item If Yes, what are the identified relationships between different human aspect(s)?
  \item Does the research include how the human aspect(s) impact on RE ? (Yes/ No)
  \item If Yes, what is the nature of the impact of the human aspect(s) on RE? (Positive/ Negative/ Both)
  \item If Positive, does the study mention the benefits of considering the aspect(s)?
  \item If Negative, how it will impact on RE?
  \item Does the study suggest any approach to mitigate the negative impact?
  \item Main outcome/ Results of the study?
  \item Does the study come up with a framework/ model as the final outcome?
  \item If Yes, what type of framework it is? (Elaborate the developed framework)
  \item How do they evaluate their results/ framework/ model?
  \item What are the major recommendations of the study?
  \item The number of citations of the study?
\end{enumerate}
\end{footnotesize}

\section{List of studies included in the SLR} \label{B}
\begin{footnotesize}
\setlength{\parindent}{4pt}
\textbf{IEE01:} E. Alatawi, A. Mendoza, and T. Miller, "Psychologically-Driven Requirements Engineering: A Case Study in Depression Care," in 2018 25th Australasian Software Engineering Conference (ASWEC), 26-30 Nov. 2018 2018, pp. 41-50, doi: 10.1109/ASWEC.2018.00014. 

\textbf{IEE02:} D. E. Damian and D. Zowghi, "The impact of stakeholders' geographical distribution on managing requirements in a multi-site organization," in Proceedings IEEE Joint International Conference on Requirements Engineering, 9-13 Sept. 2002 2002, pp. 319-328, doi: 10.1109/ICRE.2002.1048545. 

\textbf{IEE03:} A. Gregoriades, S. Jae-Eun, and A. Sutcliffe, "Human-centred requirements engineering," in Proceedings. 12th IEEE International Requirements Engineering Conference, 2004., 10-10 Sept. 2004 2004, pp. 154-163, doi: 10.1109/ICRE.2004.1335673.

\textbf{IEE04:} E. Weilemann, "A Winning Team - What Personality Has To Do With Software Engineering," in 2019 IEEE/ACM 41st International Conference on Software Engineering: Companion Proceedings (ICSE-Companion), 25-31 May 2019 2019, pp. 252-253, doi: 10.1109/ICSE-Companion.2019.00100. 

  \textbf{IEE05:} M. Mesgari, C. Okoli, and A. O. d. Guinea, "Creating Rich and Representative Personas by Discovering Affordances," IEEE Transactions on Software Engineering, vol. 45, no. 10, pp. 967-983, 2019, doi: 10.1109/TSE.2018.2826537. 

  \textbf{IEE06:} C. Solís and N. Ali, "Distributed Requirements Elicitation Using a Spatial Hypertext Wiki," in 2010 5th IEEE International Conference on Global Software Engineering, 23-26 Aug. 2010 2010, pp. 237-246, doi: 10.1109/ICGSE.2010.35. 

  \textbf{IEE07:} N. Maiden, S. Jones, K. Karlsen, R. Neill, K. Zachos, and A. Milne, "Requirements Engineering as Creative Problem Solving: A Research Agenda for Idea Finding," in 2010 18th IEEE International Requirements Engineering Conference, 27 Sept.-1 Oct. 2010 2010, pp. 57-66, doi: 10.1109/RE.2010.16.. 

  \textbf{IEE08:} D. A. Tamburri, P. Lago, H. V. Vliet, and E. d. Nitto, "On the Nature of GSE Organizational Social Structures: An Empirical Study," in 2012 IEEE Seventh International Conference on Global Software Engineering, 27-30 Aug. 2012 2012, pp. 114-123, doi: 10.1109/ICGSE.2012.25. 

  \textbf{IEE09:} S. Ludi, "Introducing Accessibility Requirements through External Stakeholder Utilization in an Undergraduate Requirements Engineering Course," in 29th International Conference on Software Engineering (ICSE'07), 20-26 May 2007 2007, pp. 736-743, doi: 10.1109/ICSE.2007.46. 

  \textbf{IEE10:} A. Sutcliffe, "Emotional requirements engineering," in 2011 IEEE 19th International Requirements Engineering Conference, 29 Aug.-2 Sept. 2011 2011, pp. 321-322, doi: 10.1109/RE.2011.6051680. 

  \textbf{IEE11:} L. Fernández-Sanz and S. Misra, "Analysis of cultural and gender influences on teamwork performance for software requirements analysis in multinational environments," IET Software, vol. 6, no. 3, pp. 167-175, 2012, doi: 10.1049/iet-sen.2011.0070.

  \textbf{IEE12:} A. Sutcliffe, "User-oriented requirements engineering," in 2014 IEEE 2nd International Workshop on Usability and Accessibility Focused Requirements Engineering (UsARE), 25-25 Aug. 2014 2014, pp. 1-8, doi: 10.1109/UsARE.2014.6890993. 

  \textbf{IEE13:} S. Thew and A. Sutcliffe, "Investigating the Role of 'Soft Issues' in the RE Process," in 2008 16th IEEE International Requirements Engineering Conference, 8-12 Sept. 2008 2008, pp. 63-66, doi: 10.1109/RE.2008.35. 

  \textbf{IEE14:} A. M. Aranda, O. Dieste, and N. Juristo, "Effect of Domain Knowledge on Elicitation Effectiveness: An Internally Replicated Controlled Experiment," IEEE Transactions on Software Engineering, vol. 42, no. 5, pp. 427-451, 2016, doi: 10.1109/TSE.2015.2494588.

  \textbf{IEE15:} G. A. Dafoulas and L. A. Macaulay, "Facilitating group formation and role allocation in software engineering groups," in Proceedings ACS/IEEE International Conference on Computer Systems and Applications, 25-29 June 2001 2001, pp. 352-359, doi: 10.1109/AICCSA.2001.934012. 

  \textbf{IEE16:} K. Manjunath, V. Anu, G. Walia, and G. Bradshaw, "Training Industry Practitioners to Investigate the Human Error Causes of Requirements Faults," in 2018 IEEE International Symposium on Software Reliability Engineering Workshops (ISSREW), 15-18 Oct. 2018 2018, pp. 53-58, doi: 10.1109/ISSREW.2018.00-31. 

  \textbf{IEE17:} A. Hoffmann and C. Lescher, "Collaboration and Intercultural Issues on Requirements:Communication, Understanding and Softskills (CIRCUS)," in 2009 Collaboration and Intercultural Issues on Requirements: Communication, Understanding and Softskills, 31-31 Aug. 2009 2009, pp. 1-4, doi: 10.1109/CIRCUS.2009.1. 

  \textbf{IEE18:} A. Nguyen-Duc, L. Jaccheri, and P. Abrahamsson, "An Empirical Study on Female Participation in Software Project Courses," in 2019 IEEE/ACM 41st International Conference on Software Engineering: Companion Proceedings (ICSE-Companion), 25-31 May 2019 2019, pp. 240-241, doi: 10.1109/ICSE-Companion.2019.00094. 

  \textbf{IEE19:} A. Hoffmann, "REIM — An improvisation workshop format to train soft skill awareness," in 2012 5th International Workshop on Co-operative and Human Aspects of Software Engineering (CHASE), 2-2 June 2012 2012, pp. 56-62, doi: 10.1109/CHASE.2012.6223023. 

  \textbf{IEE20:} J. Jung, S. Lee, S. Choi, and S. Lee, "Requirements engineering process improvement: Analyzing the organizational culture impact and implementing an empirical study to evaluate the benefits of improvement," in 2014 IEEE 1st International Workshop on the Interrelations between Requirements Engineering and Business Process Management (REBPM), 25-25 Aug. 2014 2014, pp. 15-18, doi: 10.1109/REBPM.2014.6890731. 

  \textbf{IEE21:} A. Amin, M. Rehman, S. Basri, and M. F. Hassan, "A proposed conceptual framework of programmer's creativity," in 2015 International Symposium on Technology Management and Emerging Technologies (ISTMET), 25-27 Aug. 2015 2015, pp. 108-113, doi: 10.1109/ISTMET.2015.7359011. 

  \textbf{IEE22:} A. Sutcliffe, J. Galliers, and S. Minocha, "Human errors and system requirements," in Proceedings IEEE International Symposium on Requirements Engineering (Cat. No.PR00188), 11-11 June 1999 1999, pp. 23-30, doi: 10.1109/ISRE.1999.777982. 

  \textbf{IEE23:} J. Ralyté, "Viewpoints and Issues in Requirements Engineering for Services," in 2012 IEEE 36th Annual Computer Software and Applications Conference Workshops, 16-20 July 2012 2012, pp. 341-346, doi: 10.1109/COMPSACW.2012.68. 

  \textbf{IEE24:} T. Alsanoosy, M. Spichkova, and J. Harland, "Cultural Influences on the Requirements Engineering Process: Lessons Learned from Practice," in 2018 23rd International Conference on Engineering of Complex Computer Systems (ICECCS), 12-14 Dec. 2018 2018, pp. 61-70, doi: 10.1109/ICECCS2018.2018.00015. 

  \textbf{IEE25:} P. K. Murukannaiah, N. Ajmeri, and M. P. Singh, "Acquiring Creative Requirements from the Crowd: Understanding the Influences of Personality and Creative Potential in Crowd RE," in 2016 IEEE 24th International Requirements Engineering Conference (RE), 12-16 Sept. 2016 2016, pp. 176-185, doi: 10.1109/RE.2016.68. 

  \textbf{IEE26:} M. Z. H. Kolpondinos and M. Glinz, "Tailoring Gamification to Requirements Elicitation: A Stakeholder-Centric Motivation Concept," in 2017 IEEE/ACM 10th International Workshop on Cooperative and Human Aspects of Software Engineering (CHASE), 23-23 May 2017 2017, pp. 9-15, doi: 10.1109/CHASE.2017.4. 

  \textbf{IEE27:} A. Marnewick, J. Pretorius, and L. Pretorius, "A perspective on human aspects contributing to quality requirements: A cross-case analysis," in 2011 IEEE International Conference on Industrial Engineering and Engineering Management, 6-9 Dec. 2011 2011, pp. 389-393, doi: 10.1109/IEEM.2011.6117945 

  \textbf{ACM01:} E. M. Trauth, J. L. Quesenberry, and B. Yeo, Environmental influences on gender in the IT workforce (no. 1). Association for Computing Machinery, 2008, pp. 8–32. 

  \textbf{ACM02:} L. F. Capretz and F. Ahmed, Why do we need personality diversity in software engineering? (no. 2). Association for Computing Machinery, 2010, pp. 1–11. 

  \textbf{ACM03:} J. Verner, S. Beecham, and N. Cerpa, Stakeholder dissonance: disagreements on project outcome and its impact on team motivation across three countries (Proceedings of the 2010 Special Interest Group on Management Information System’s 48th annual conference on Computer personnel research on Computer personnel research). Vancouver, BC, Canada: Association for Computing Machinery, 2010, pp. 25–33. 

  \textbf{ACM04:} L. G. Martínez, G. Licea, A. Rodríguez-Díaz, and J. R. Castro, Experiences in software engineering courses using psychometrics with RAMSET (Proceedings of the fifteenth annual conference on Innovation and technology in computer science education). Bilkent, Ankara, Turkey: Association for Computing Machinery, 2010, pp. 244–248. 

  \textbf{ACM05:} A. C. C. França and F. Q. B. d. Silva, Designing motivation strategies for software engineering teams: an empirical study (Proceedings of the 2010 ICSE Workshop on Cooperative and Human Aspects of Software Engineering). Cape Town, South Africa: Association for Computing Machinery, 2010, pp. 84–91. 

  \textbf{ACM06:} C. França, H. Sharp, and F. Q. B. d. Silva, Motivated software engineers are engaged and focused, while satisfied ones are happy (Proceedings of the 8th ACM/IEEE International Symposium on Empirical Software Engineering and Measurement). Torino, Italy: Association for Computing Machinery, 2014, p. Article 32. 

  \textbf{ACM07:} M. E. Hoffman, P. V. Anderson, and M. Gustafsson, Workplace scenarios to integrate communication skills and content: a case study (Proceedings of the 45th ACM technical symposium on Computer science education). Atlanta, Georgia, USA: Association for Computing Machinery, 2014, pp. 349–354. 

  \textbf{ACM08:} G. Calikli, M. Al-Eryani, E. Baldebo, J. Horkofff, and A. Ask, Effects of automated competency evaluation on software engineers’ emotions and motivation: a case study (Proceedings of the 3rd International Workshop on Emotion Awareness in Software Engineering). Gothenburg, Sweden: Association for Computing Machinery, 2018, pp. 44–50. 

  \textbf{ACM09:} A. D. Fucci et al., Needs and challenges for a platform to support large-scale requirements engineering: a multiple-case study (Proceedings of the 12th ACM/IEEE International Symposium on Empirical Software Engineering and Measurement). Oulu, Finland: Association for Computing Machinery, 2018, p. Article 19. 

  \textbf{ACM10:} S. Seibel and N. Veilleux, Factors influencing women entering the software development field through coding bootcamps vs. computer science bachelor’s degrees (no. 6). Consortium for Computing Sciences in Colleges, 2019, pp. 84–96. 

  \textbf{ACM11:} C. Wang, P. Cui, M. Daneva, and M. Kassab, Understanding what industry wants from requirements engineers: an exploration of RE jobs in Canada (Proceedings of the 12th ACM/IEEE International Symposium on Empirical Software Engineering and Measurement). Oulu, Finland: Association for Computing Machinery, 2018, p. Article 41. 

  \textbf{ACM12:} D. V. Pereira, G. M. Corrêa, F. Q. B. d. Silva, and D. M. Ribeiro, Team maturity in software engineering teams: a work in progress (Proceedings of the 10th International Workshop on Cooperative and Human Aspects of Software Engineering). Buenos Aires, Argentina: IEEE Press, 2017, pp. 70–73. 

  \textbf{ACM13:} E. Winter, S. Forshaw, L. Hunt, and M. A. Ferrario, Advancing the study of human values in software engineering (Proceedings of the 12th International Workshop on Cooperative and Human Aspects of Software Engineering). Montreal, Quebec, Canada: IEEE Press, 2019, pp. 19–26. 

  \textbf{ACM14:} D. Ford, T. Zimmermann, C. Bird, and N. Nagappan, Characterizing software engineering work with personas based on knowledge worker actions (Proceedings of the 11th ACM/IEEE International Symposium on Empirical Software Engineering and Measurement). Markham, Ontario, Canada: IEEE Press, 2017, pp. 394–403. 

  \textbf{ACM15:} L. F. Capretz, D. Varona, and A. Raza, Influence of personality types in software tasks choices (no. C). Elsevier Science Publishers B. V., 2015, pp. 373–378.  

  \textbf{ACM16:} Y. Wang and D. Redmiles, Implicit gender biases in professional software development: an empirical study (Proceedings of the 41st International Conference on Software Engineering: Software Engineering in Society). Montreal, Quebec, Canada: IEEE Press, 2019, pp. 1–10. 

  \textbf{ACM18:} R. Colomo-Palacios, T. Samuelsen, and C. Casado-Lumbreras, Emotions in software practice: presentation vs. coding (Proceedings of the 4th International Workshop on Emotion Awareness in Software Engineering). Montreal, Quebec, Canada: IEEE Press, 2019, pp. 23–28.

  \textbf{ACM19:} T. Miller, S. Pedell, A. A. Lopez-Lorca, A. Mendoza, L. Sterling, and A. Keirnan, Emotion-led modelling for people-oriented requirements engineering (no. C). Elsevier Science Inc., 2015, pp. 54–71. 

  \textbf{ACM20:} F. Anvari, D. Richards, M. Hitchens, M. A. Babar, H. M. T. Tran, and P. Busch, An empirical investigation of the influence of persona with personality traits on conceptual design (no. C). Elsevier Science Inc., 2017, pp. 324–339.   

  \textbf{SP01:} N. C. Pa and A. M. Zin, "Managing Communications Challenges in Requirement Elicitation," in Software Engineering and Computer Systems, Berlin, Heidelberg, J. M. Zain, W. M. b. Wan Mohd, and E. El-Qawasmeh, Eds., 2011, pp. 803-811.    

  \textbf{SP02:} F. Anwar, R. Razali, and K. Ahmad, "Achieving Effective Communication during Requirements Elicitation - A Conceptual Framework," in Software Engineering and Computer Systems, Berlin, Heidelberg, J. M. Zain, W. M. b. Wan Mohd, and E. El-Qawasmeh, Eds., 2011, pp. 600-610.   
 
  \textbf{SP03:} N. Condori-Fernandez, S. España, K. Sikkel, M. Daneva, and A. González, "Analyzing the Effect of the Collaborative Interactions on Performance of Requirements Validation," in Requirements Engineering: Foundation for Software Quality, Cham, C. Salinesi and I. van de Weerd, Eds., 2014, pp. 216-231.   
  
  \textbf{SP04:} M. Mahaux, A. Mavin, and P. Heymans, "Choose Your Creativity: Why and How Creativity in Requirements Engineering Means Different Things to Different People," in Requirements Engineering: Foundation for Software Quality, Berlin, Heidelberg, B. Regnell and D. Damian, Eds., 2012, pp. 101-116.  

  \textbf{SP05:} N. K. Sethia and A. S. Pillai, "The Effects of Requirements Elicitation Issues on Software Project Performance: An Empirical Analysis," in Requirements Engineering: Foundation for Software Quality, Cham, C. Salinesi and I. van de Weerd, Eds., 2014, pp. 285-300.  

  \textbf{SP06:} H. Bendjenna, N. Zarour, and P.-J. Charrel, "Enhancing Elicitation Technique Selection Process in a Cooperative Distributed Environment," in Requirements Engineering: Foundation for Software Quality, Berlin, Heidelberg, B. Paech and C. Rolland, Eds., 2008, pp. 23-36.    

  \textbf{SP07:} S. De Ascaniis, L. Cantoni, E. Sutinen, and R. Talling, "A LifeLike Experience to Train User Requirements Elicitation Skills," in Design, User Experience, and Usability: Understanding Users and Contexts, Cham, A. Marcus and W. Wang, Eds., 2017, pp. 219-237.  
  
  \textbf{SP08:} T. Tuunanen, K. Peffers, and S. Hebler, "A Requirements Engineering Method Designed for the Blind," in Global Perspectives on Design Science Research, Berlin, Heidelberg, R. Winter, J. L. Zhao, and S. Aier, Eds., 2010, pp. 475-489.  

  \textbf{SP09:} C. L. de la Barra, B. Crawford, R. Soto, S. Misra, and E. Monfroy, "Agile Software Development: It Is about Knowledge Management and Creativity," in Computational Science and Its Applications – ICCSA 2013, Berlin, Heidelberg, B. Murgante et al., Eds., 2013, pp. 98-113.  
 
  \textbf{SP10:} M. Niazi and M. A. Babar, "De-motivators of Software Process Improvement: An Analysis of Vietnamese Practitioners’ Views," in Product-Focused Software Process Improvement, Berlin, Heidelberg, J. Münch and P. Abrahamsson, Eds., 2007, pp. 118-131.  

  \textbf{SP11:} F. Anvari and D. Richards, "A Method to Identify Talented Aspiring Designers in Use of Personas with Personality," in Evaluation of Novel Approaches to Software Engineering, Cham, L. A. Maciaszek and J. Filipe, Eds., 2016, pp. 40-61.  
 
  \textbf{SP12:} L. Bormane and S. Berzisa, "Role of “Bridge Person” in Software Development Projects," in Information and Software Technologies, Cham, R. Damaševičius and V. Mikašytė, Eds., 2017, pp. 3-14.  

  \textbf{SP13:} L. Fernandez-Sanz and S. Misra, "Influence of human factors in Software Quality and Productivity," in Computational Science and Its Applications - ICCSA 2011, Berlin, Heidelberg, B. Murgante, O. Gervasi, A. Iglesias, D. Taniar, and B. O. Apduhan, Eds., 2011, pp. 257-269.  
 
  \textbf{SP15:} J. Griffyth, "Human Factors in High Integrity Software Development: a Field Study," in Safe Comp 96, London, E. Schoitsch, Ed., 1997, pp. 301-310.   
 
  \textbf{SP16:} S. Dorairaj, J. Noble, and P. Malik, "Effective Communication in Distributed Agile Software Development Teams," in Agile Processes in Software Engineering and Extreme Programming, Berlin, Heidelberg, A. Sillitti, O. Hazzan, E. Bache, and X. Albaladejo, Eds., 2011, pp. 102-116.  

  \textbf{SP17:} C. J. Stettina and W. Heijstek, "Five Agile Factors: Helping Self-management to Self-reflect," in Systems, Software and Service Process Improvement, Berlin, Heidelberg, R. V. O‘Connor, J. Pries-Heje, and R. Messnarz, Eds., 2011, pp. 84-96.   

  \textbf{WI01:} N. Baddoo, T. Hall, and C. O'Keeffe, "Using multi dimensional scaling to analyse software engineers' de-motivators for SPI," Software Process: Improvement and Practice, vol. 12, no. 6, pp. 511-522, 2007, doi: 10.1002/spip.352.  
 
  \textbf{WI02:} L. Freund, E. G. Toms, and J. Waterhouse, "Modeling the information behaviour of software engineers using a work - task framework," Proceedings of the American Society for Information Science and Technology, vol. 42, no. 1, 2005, doi: 10.1002/meet.14504201181.  

  \textbf{WI03:} D. Graziotin, X. Wang, and P. Abrahamsson, "Do feelings matter? On the correlation of affects and the self-assessed productivity in software engineering," Journal of Software: Evolution and Process, vol. 27, no. 7, pp. 467-487, 2015, doi: 10.1002/smr.1673.  
 
  \textbf{WI04:} R. Klendauer, M. Berkovich, R. Gelvin, J. M. Leimeister, and H. Krcmar, "Towards a competency model for requirements analysts," Information Systems Journal, vol. 22, no. 6, pp. 475-503, 2012, doi: 10.1111/j.1365-2575.2011.00395.x.  
 
  \textbf{SB01:} S. T. Acuña, M. Gómez, and N. Juristo, "How do personality, team processes and task characteristics relate to job satisfaction and software quality?," Information and Software Technology, vol. 51, no. 3, pp. 627-639, 2009/03/01/ 2009, doi: https://doi.org/10.1016/j.infsof.2008.08.006.  
 
  \textbf{SB02:} C. J. Davis, R. M. Fuller, M. C. Tremblay, and D. J. Berndt, "Communication Challenges in Requirements Elicitation and the Use of the Repertory Grid Technique," Journal of Computer Information Systems, vol. 46, no. 5, pp. 78-86, 2006/06/01 2006, doi: 10.1080/08874417.2006.11645926.  

  \textbf{SB03:} R. Feldt, R. Torkar, L. Angelis, and M. Samuelsson, "Towards individualized software engineering: empirical studies should collect psychometrics," presented at the Proceedings of the 2008 international workshop on Cooperative and human aspects of software engineering, Leipzig, Germany, 2008. [Online]. Available: https://doi.org/10.1145/1370114.1370127.  
  
  \textbf{SB04:} J. S. Karn and A. J. Cowling, "Using ethnographic methods to carry out human factors research in software engineering," Behavior Research Methods, vol. 38, no. 3, pp. 495-503, 2006/08/01 2006, doi: 10.3758/BF03192804.  

  \textbf{SB05:} S. Thew and A. Sutcliffe, "Value-based requirements engineering: method and experience," Requirements Engineering, vol. 23, no. 4, pp. 443-464, 2018/11/01 2018, doi: 10.1007/s00766-017-0273-y.  
   
  \textbf{SB06:} M. V. Kosti, R. Feldt, and L. Angelis, "Personality, emotional intelligence and work preferences in software engineering: An empirical study," Information and Software Technology, vol. 56, no. 8, pp. 973-990, 2014/08/01/ 2014, doi: https://doi.org/10.1016/j.infsof.2014.03.004.    
   
  \textbf{SB07:} G. Aranda and M. Piattini, "A framework to improve communication during the requirements elicitation process in GSD projects," Requir. Eng., vol. 15, pp. 397-417, 11/01 2010, doi: 10.1007/s00766-010-0105-9.  
  
  \textbf{SB08:} Holtkamp, P., Jokinen, J. P., \& Pawlowski, J. M. (2015). Soft competency requirements in requirements engineering, software design, implementation, and testing. Journal of Systems and Software, 101, 136-146.
 
 \end{footnotesize}



%

\bibliographystyle{ieeetr}
\bibliography{output}

%

\vspace{-1cm}
\begin{IEEEbiography}[{\includegraphics[width=1in,height=1.25in,clip,keepaspectratio]{Dulaji.pdf}}] {Dulaji Hidellaarachchi} is a PhD candidate at Monash University, Melbourne, Australia. She received her BSc (Hons.) in Information Systems degree from University of Colombo School of Computing (UCSC), Sri Lanka. Prior to her PhD candidature, she was in academia as an assistant lecturer and an instructor. Her research interests are requirements engineering, human and social aspects of software engineering, human computer interaction and information systems design. More details of her research can be found at, https://www.researchgate.net/
profile/DulajiHidellaarachchi. Contact her at:  dulaji.hidellaarachchi@monash.edu

\end{IEEEbiography}

\vspace{-1cm}

\begin{IEEEbiography}[{\includegraphics[width=1in,height=1.25in,clip,keepaspectratio]{john.pdf}}]{John Grundy}received the BSc (Hons), MSc, and PhD degrees in computer science from the University of Auckland, New Zealand. He is an Australian Laureate fellow and a professor of software engineering at Monash University, Melbourne, Australia. He is an associate editor of the IEEE Transactions on Software Engineering, the Automated Software Engineering Journal, and IEEE Software. His current interests include domain--specific visual languages, model--driven engineering, large-scale systems engineering, and software engineering education. More details about his research can be found at https://sites.google.com/site/johncgrundy/. Contact him at john.grundy@monash.edu.

\end{IEEEbiography}

\vspace{-1cm}

\begin{IEEEbiography}[{\includegraphics[width=1in,height=1.25in,clip,keepaspectratio]{rashina.pdf}}]{Rashina Hoda}is an associate professor in software engineering at Monash University, Melbourne, Australia. She received her PhD in computer science from Victoria University of Wellington, New Zealand. Her areas of research interests include human and social aspects of software engineering, grounded theory, and serious game design. She serves on the IEEE Transactions on Software Engineering review board, IEEE Software Advisory Board, as ICSE2021 social media co--chair, CHASE 2021 program co--chair, and XP2020 program co--chair. More details about her research can be found at https://rashina.com. Contact her at rashina.hoda@monash.edu.

\end{IEEEbiography}

\vspace{-1cm} 
\begin{IEEEbiography}[{\includegraphics[width=1in,height=1.25in,clip,keepaspectratio]{kashumi.pdf}}]{Kashumi Madampe}is a Graduate Student Member of IEEE and is with Monash University, Melbourne, Australia. Ms. Madampe did part of her current research at The University of Auckland, New Zealand. Prior to the PhD candidature, she was in the software development industry as a project manager and a business analyst. Her research interests are requirements engineering, human and social aspects of software engineering, software repository mining, grounded theory, and natural language processing. She serves as the ASE2021 publicity and social media chair, CHASE2021 social media chair, and XP2021 Poster co--chair. More details about her research can be found at https://kashumim.com. Contact her at kashumi.madampe@monash.edu.
\end{IEEEbiography}







\end{document}